\documentclass[12pt]{article}
\usepackage{amsmath}
\usepackage{graphicx,psfrag,epsf,amssymb,amsxtra,mathtools}
\usepackage{enumerate}

\usepackage{amsfonts}

\usepackage{url} 
\usepackage{slashbox}

\usepackage{bbm}
\usepackage{dsfont}
\usepackage{booktabs}
\usepackage{xcolor}
\usepackage[multiple]{footmisc}

\usepackage{natbib}

\usepackage[hidelinks]{hyperref}
\usepackage[absolute,overlay]{textpos} 

\newcommand{\jbesattribution}{%
  \noindent\small This is an Accepted Manuscript of an article published by Taylor \& Francis in the \textit{Journal of Business \& Economic Statistics} on 26 Feb 2024, available online: \url{https://doi.org/10.1080/07350015.2024.2308121}%
}

\newcommand{\blind}{0}

\newcommand{\Ho}{\mathrm{H}_0}

\newcommand{\betabar}{\bar{\beta}}
\newcommand{\betahat}{\hat{\beta}}

\newcommand{\1}{\mathbbm{1}}
\newcommand{\E}{\mathbb{E}}

\newcommand{\N}{\mathbb{N}}
\newcommand{\R}{\mathbb{R}}

\newcommand{\calU}{\mathcal{U}}

\newcommand{\defined}{\vcentcolon=}

\newcommand{\plim}{\operatorname{plim}}

\newcommand{\iid}{\mathrm{i.i.d.}}

\newcommand{\betahatTplusone}{\hat{\pi}_{ATT}}

\newcommand{\calR}{\mathcal{R}}

\newcommand{\dto}{\overset{d}{\to}}
\newcommand{\pto}{\overset{p}{\to}}

\newcommand{\Yitdots}{\ddot{Y}_{i,t}}
\newcommand{\Witdots}{\ddot{W}_{i,t}}
\newcommand{\uitdots}{\ddot{u}_{i,t}}
\newcommand{\Ybaridot}{\bar{Y}_{i,\cdot}}
\newcommand{\Ybardott}{\bar{Y}_{\cdot, t}}
\newcommand{\Ybardotdot}{\bar{Y}_{\cdot\cdot}}

\newcommand{\Yidots}{\ddot{Y}_{i}}
\newcommand{\Widots}{\ddot{W}_{i}}
\newcommand{\uidots}{\ddot{u}_{i}}

\newcommand{\sumi}{\sum_{i=1}^n}

\newcommand{\uit}{u_{i,t}}

\newcommand{\Yit}{Y_{i,t}}

\newcommand{\Gi}{\mathbb{G}_i}

\newtheorem{thm}{Theorem}[section]

\newtheorem{ass}{Assumption}[section]

\newtheorem{rem}{Remark}[section]

\numberwithin{equation}{section}

\begin{document}

\def\spacingset#1{\renewcommand{\baselinestretch}%
{#1}\small\normalsize} \spacingset{1}


\if0\blind
{
  \title{\bf Testing for equivalence of pre-trends in Difference-in-Differences estimation}
  \author{Holger Dette 
    \\
    Department of Mathematics, Ruhr University Bochum\\
    and \\
    Martin Schumann \\
    School of Business and Economics, Maastricht University \thanks{\jbesattribution}}
  \maketitle
} \fi

\if1\blind
{
  \bigskip
  \bigskip
  \bigskip
  \begin{center}
    {\LARGE\bf Testing for equivalence of pre-trends in Difference-in-Differences estimation}
\end{center}
  \medskip
} \fi

\bigskip
\begin{abstract}
\noindent The plausibility of the ``parallel trends assumption'' in Difference-in-Differences estimation is usually assessed by a test of the null hypothesis that the difference  between the average outcomes of both groups is constant over time before the treatment. However, failure to reject the null hypothesis does not imply the absence of differences in time trends between both groups. We provide equivalence tests that allow researchers to find evidence in favor of the parallel trends assumption and thus increase the credibility of their treatment effect estimates.  While we motivate our tests in the standard two-way fixed effects model, we discuss simple extensions to settings in which treatment adoption is staggered over time.
\end{abstract}

\noindent%
\vfill

\newpage
\spacingset{1.45} 
\section{Introduction}

In the classic case, the Difference-in-Differences (DiD) framework consists of two groups observed over two periods of time, where the ``treatment group'' is untreated in the initial period and has received a treatment in the second period whereas the ``control group'' is untreated in both periods. The key condition under which the DiD estimator yields sensible point estimates of the average treatment effect on the treated is known as the ``parallel paths'' or ``parallel trends assumption'', henceforth referred to as PTA, which states that in the absence of treatment both groups would have experienced the same temporal trends in the outcome variable on average. If pre-treatment observations are available for both groups, the plausibility of this assumption is typically assessed by plots accompanied by a formal testing procedure showing that there is no evidence of differences in trends over time between the treatment and the control group. However, traditional pre-tests can suffer from low power to detect violations of the PTA \citep{kahn2020promise}. Thus, finding no evidence of differences in trends in finite samples does not imply that there are no differences in trends in the population. More concerningly, \cite{roth2018pre} points out that if differences in trends exist, conditional on not detecting violations of parallel trends at the pre-testing stage, the bias of DiD-estimators may be greatly amplified. 

Given the severe consequences of falsely accepting the PTA, we propose that instead of testing the null hypothesis of "no differences in trends" between the treatment and the control group in the pre-treatments periods, one should apply a test for statistical equivalence. We provide three distinct types of equivalence that impose bounds on the maximum, the average and the root mean square change over time in the group mean difference between treatment and control in the pre-treatment periods. Given a threshold below which deviations from the PTA can be considered negligible, these tests allow the researcher to provide statistical evidence in favor of the PTA, thus increasing its credibility. If no sensible equivalence threshold can be determined before analyzing the data, we propose to report the smallest equivalence threshold for which the null hypothesis of "non-negligible trend differences" can still be rejected at a given level of significance. Conceptually, this idea is similar to the ``equivalence confidence interval'' in \cite{hartman2018} applied to a DiD setting. Our procedure reverses the burden of proof since the data has to provide evidence \textit{in favor} of similar trends in the treatment and the control group, which is arguably more appropriate for an assumption as crucial to the DiD-framework as the PTA. Furthermore, the power to reject the null hypothesis of ``non-negligible differences'' is increasing with the sample size (also see \citealp{hartman2018}). This improves upon the current practice of testing the null hypothesis of ``no difference'', since large samples increase the chances of rejecting this null hypothesis (and thus seemingly making the DiD framework inapplicable), even if the true difference between treatment and control may be negligible in the given context. Finally, our equivalence test statistics can easily be implemented in practice. While we motivate our tests in the standard two-way fixed effects model with panel data, we discuss how our tests can be applied in situations where treatment timing differs across groups (i.e.\ ``staggered treatment assignment'') or where average treatment effects depend on some observable characteristics. 

As we use equivalence tests, our paper is closely related to \cite{bilinski2020nothing}, who provide a discussion on the benefits of using equivalence (or "non-inferiority") tests when testing for violations of modeling assumptions. Their ``one-step-up'' approach is based on a non-inferiority test of treatment effect estimates obtained from a standard DiD model and from a model augmented with a particular violation of the parallel trends assumption (e.g.\ a linear trend). While both approaches stress the potential benefits of equivalence testing in DiD setups, a distinctive feature is that we do not necessarily focus on a particular violation of the PTA. As pointed out in \cite{kahn2020promise}, including for instance group-specific linear time trends can lead to a loss in degrees of freedom and thus to a substantial loss in power. In contrast, our approach focuses on testing for any non-negligible differences between treatment and control in the pre-treatment periods. Our paper is also related to other approaches that allow for certain deviations from exactly parallel trends. 
\cite{roth2020honest} partially identify the ATT under restrictions that impose that the post-treatment violation of parallel trends is not too large relative to the pre-treatment violation. By contrast, we propose tests for the null hypothesis that the pre-treatment violation of parallel trends is large. If the null is rejected, so that the pre-treatment violation is determined to be small, then the researcher may decide that violations of parallel trends can safely be ignored with minimal bias. Alternatively, the researcher could use the upper bound on the pre-treatment violation given by our approach as a way of determining reasonable bounds on the post-treatment violations of parallel trends, which could then be used as input to the partial identification frameworks of \cite{roth2020honest} or \cite{manski2018right}. 

The rest of the paper is organized as follows. Section 2 introduces the main two-way fixed effects model and discusses the widely used practice of testing for violations of the PTA. Equivalence tests and our hypotheses are discussed in Section 3. Section 4 introduces our assumptions and presents the test statistics for our hypotheses. In Section 5, we present extensions of the main model that allow for heterogeneous treatment effects due to differences in treatment timing or observable characteristics. Simulation evidence on the performance of our test procedures is provided in Section 6, while Section 7 contains an empirical illustration of our approach. Section 8 concludes. Finally, mathematical details and tables are collected in the Appendix.
\section{Pre-testing in the canonical DiD model}\label{section pre-testing in the canonical}
We initially contrast our test procedures with the usual pre-test in the canonical DiD case with only two groups and common treatment timing. The researcher observes a balanced panel of $n$ individuals recorded over $T+2$ periods of time. We assume that in the first $T+1$ periods none of the individuals is treated whereas in period $T+2$ a subset has received treatment. Individual $i$ is a member of the treatment group if $G_i=1$ and a member of the control group if $G_i=0$. Following \cite{kahn2020promise} we refer to period $T+1$ as the ``base period'' as treatment effects are usually assessed by comparing post-treatment outcomes with the outcomes in the last pre-treatment period.  The potential outcomes of unit $i$ in period $t$ when treated and in the absence of treatment are denoted as $Y_{i,t}(1)$ and $Y_{i,t}(0)$, respectively.
The object of interest is the average treatment effect on the treated (ATT), given as
\begin{equation*}
\pi_{ATT}\defined \E[Y_{i,T+2}(1)-Y_{i,T+2}(0)|G_i=1].
\end{equation*}
For identification of the ATT, we need to make several assumptions. First, we require that $\Pr(G_i=1)=p\in (0,1)$, which is  an ``overlap'' condition that ensures that both treatment and control are non-empty (\citealp{santanna-zhao-2020}). Next, we assume "no-anticipation", which rules out treatment effects before the actual treatment date. Adapting \cite{borusyak}, we assume
\begin{equation}\label{no-anticipation new}
\E[Y_{i,t}|G_i]=\E[Y_{i,t}(0)|G_i]+\pi_{ATT}G_iD_{T+2}(t),~ t=1,...,T+2,    
\end{equation}
with $D_{l}(t) =\1\{l=t\}$. This implies that the expected observed outcome coincides with the expected potential outcome corresponding to the actual treatment status both for the control units and the eventually treated units. Next, let
\begin{equation}\label{Y0 model}
\E[Y_{i,t}(0)|G_i]=\alpha_i+\lambda_t + G_i\gamma_t,    
\end{equation}
where $\alpha_i, \lambda_t$ and $\gamma_t$ are some non-stochastic constants. Combining \eqref{no-anticipation new} with \eqref{Y0 model}, we can write
\begin{equation}\label{model intermediate}
    Y_{i,t}=\alpha_i+\lambda_t + G_i\gamma_t+\pi_{ATT}G_iD_{T+2}(t)+u_{i,t},
\end{equation}
with $u_{i,t}=Y_{i,t}-\E[Y_{i,t}|G_i]$. It is clear from \eqref{model intermediate} that $\pi_{ATT}$ is not identified without further restrictions on $\gamma_t$. The fundamental assumption that leads to the DiD estimator is the (augmented) PTA
\begin{equation}\label{statement parallel trends new}
\E[Y_{i,t}(0)-Y_{i,T+1}(0)|G_i=1]-\E[Y_{i,t}(0)-Y_{i,T+1}(0)|G_i=0]=0,~ t=1,...,T+2,
\end{equation}	
which implies that $\gamma_t-\gamma_{T+1}=0$ for all $t=1,...,T+2$. Notice that this assumption over-identifies $\pi_{ATT}$, as identification only requires parallel trends between the post-treatment and the base period, i.e.\ $\gamma_{T+2}-\gamma_{T+1}=0$. However, as it is often difficult to imagine circumstances under which the latter condition is satisfied whereas \eqref{statement parallel trends new} is not, the augmented PTA can be useful for a pre-testing procedure that allows researchers to assess the plausibility of the PTA post-treatment. To do so, one uses the data from periods $1,...,T+1$ to estimate the two-way fixed effects (TWFE) model 
\begin{equation}\label{main model new}
    Y_{i,t}=\alpha_i+\lambda_t + \sum_{l=1}^{T}\beta_lG_i D_{l}(t) +u_{i,t},
\end{equation}
where $\beta_l=\gamma_l-\gamma_{T+1}$.  Here, $\beta=(\beta_1,...,\beta_T)'$ collects all ``placebo'' treatment effects so that under \eqref{no-anticipation new} and \eqref{Y0 model}, $\beta=0$ if and only if the augmented PTA holds. As shown, for instance, in \cite{Baltagi2021} or \cite{wooldridge2021two}, $\beta$ can be estimated by pooled OLS on ``double-demeaned'' data. Let $W_{i,t,l}=G_i D_l(t)$ and $W_{i,t}=(W_{i,t,1},...,W_{i,t,T})'$. Double-demeaning \eqref{main model new} then yields
\begin{equation}
\Yitdots=\Witdots'\beta+\uitdots,
\label{main model demeaned}
\end{equation}
for $ i=1,...,n$ and $t=1,...,T+1$, where $\Yitdots=\Yit-\Ybaridot-\Ybardott+\Ybardotdot$, 
\[
\Ybaridot=\frac{1}{T+1}\sum_{t=1}^{T+1}\Yit,~ \Ybardott=\frac{1}{n}\sumi\Yit,~ \Ybardotdot=\frac{1}{n(T+1)}\sumi\sum_{t=1}^{T+1}\Yit
\]
and $\Witdots$ and $\uitdots$ are defined analogously. Since $\E[u_{i,t}|G_i]=0$ by construction, consistency and asymptotic normality of the pooled OLS estimator in \eqref{main model demeaned}, denoted as $\betahat$, follow under mild conditions. To find evidence \textit{against} the plausibility of parallel trends, it is therefore common in applied economic research to test for individual significance (see \citealp{roth2018pre}), i.e.\ for every $l\in\{1,...,T\}$ we test
\begin{equation}
\mathrm{H}_0:~\beta_l=0\qquad \mbox{vs.} \qquad\mathrm{H}_1: \beta_l\neq 0.
                        \label{individual H0}
\end{equation}
If the null hypothesis is rejected in a pre-treatment period, the PTA is deemed unreasonable, and consequently the DiD framework is often regarded as unsuitable in the corresponding context. This procedure has several shortcomings. For instance, a problematic common practice is to treat failure to reject the null hypothesis in  \eqref{individual H0} as evidence \textit{in favor} of $\mathrm{H}_0$, i.e.\ one proceeds as if the null hypothesis was true and as if the PTA held. From a statistical point of view, this practice is incorrect as it neglects the error of type II. In some cases, there may be differences in trends between both groups in the population that cannot be detected with traditional test of  \eqref{individual H0} due to a lack of statistical power. \cite{roth2018pre} points out that ignoring these differences can amplify the bias and thus raise concerns of a ``publication bias'', since articles using a DiD identification argument are more likely to be deemed publishable when a test of \eqref{individual H0} could not detect evidence against the PTA. Moreover, the DiD framework is sometimes used even when $\mathrm{H}_0$ in \eqref{individual H0} is rejected, as some statistically significant differences are deemed negligible in a given context. However, a potential threshold $\mathcal{U}>0$ that quantifies what constitutes a negligible difference is often not adequately discussed. On the other hand, if the DiD framework is not applied when $\Ho$ in \eqref{individual H0} is rejected in at least one pre-treatment period, useful information may be lost if $\mathcal{U}$ can be interpreted as a plausible ``upper bound'' for trend differences. 
For these reasons, the plausibility of the PTA as the fundamental modeling assumption of the DiD framework can be more convincingly assessed using statistical equivalence tests as these tests address all of the above shortcomings of the current standard testing procedure. For instance, to rewrite \eqref{individual H0} in terms of statistical equivalence, for some $l\in\{1,...,T\}$ one would define the equivalence threshold $\mathcal{U}>0$ and test
\begin{equation}\label{equivalence individual H0}
\mathrm{H}_0:~|\beta_l|\geq \mathcal{U} \qquad \mbox{vs.} \qquad\mathrm{H}_1: |\beta_l|<\mathcal{U}
\end{equation}
so that rejecting $\Ho$ yields evidence \textit{in favor} of negligible trend differences between periods $l$ and $T+1$. While the hypothesis \eqref{equivalence individual H0} can easily be tested with the "two one-sided tests" procedure of \cite{schuirmann1987}, we do not recommend this approach as it would lead to an accumulation of Type I error due to multiple testing. In the following, we elaborate on the benefits of equivalence tests and provide ways of summarizing the statistical evidence in favor of the PTA in the pre-treatment periods by formulating joint hypotheses.

\section{Testing for equivalence}
Equivalence testing is well known in biostatistics (see \citealp{berger1996bioequivalence} or \citealp{wellek2010}). While it has recently been considered in the statistical literature for the analysis of structural breaks (e.g.\ \citealp{Dette2014b}, \citealp{dettewu2019}, \citealp{detkokaue2020} or \citealp{DKV2018}), it is less frequently used in econometrics.  Instead of assuming that treatment and control are perfectly comparable unless there is strong evidence \textit{against} this assumption, we suggest several testing procedures that explicitly require finding evidence \textit{in favor} of the comparability of both groups. Each of the tests is based on an upper bound $\calU> 0$ for changes in the group mean differences in the pre-treatment periods relative to the base period. There are two ways in which one can make use of the upper bound $\calU$. First, as in the ``classic'' use of equivalence tests, one can specify a threshold $\mathcal{U}$ below which changes in the group mean differences over time are deemed negligible. Rejecting the null hypothesis that trend differences are larger than $\calU$ at level of significance $\alpha$ then implies that deviations from parallel trends in the pre-treatment periods are negligible at confidence level $1-\alpha$.  The researcher may then interpret this as support for negligible differences in trends post-treatment and hence assume that the PTA holds. This procedure improves upon the current pre-test as it requires an explicit rationalization of the threshold $\mathcal{U}$ and sufficient data to support the assumption of negligible violations of the PTA. The choice of the threshold $\mathcal{U}$ should thus reflect the specific scientific background of the application. In bio-statistics, the popularity of equivalence tests has led to a consensus on sensible choices for $\mathcal{U}$, and regulators frequently specify the equivalence thresholds that should be employed (see \citealp{wellek2010} for a recent review). We expect that with a more frequent adoption of equivalence testing in applied economics a similar consensus will be reached. However, in some applications, it may still be difficult to objectively argue that a certain extent of violations of the PTA can be ignored in practice. It may then be sensible to report $\calU^*$ as the smallest value at which $\Ho$ can be rejected at a given level of significance (i.e.\ for which ``equivalence of pre-trends" can be concluded). Small values of $\calU^*$ relative to the estimated treatment effect may then be regarded as reassuring as it is unlikely that the treatment effect is merely an artifact of differences in trends. On the contrary, if $\calU^*$ is relatively large, the credibility of the estimated effect is in serious doubt. A similar idea has been proposed in \cite{hartman2018}. It can further be related to the "breakdown frontier" proposed by \cite{Masten2020}, as treatment effects can be considered non-robust to violations of the PTA when $\calU^*$ exceeds the estimated treatment effect. Finally, in cases where the choice of the threshold is difficult, the methodology presented here can also be used to provide (asymptotic) confidence intervals for violations of the PTA. 
\subsection{Formulating equivalence hypotheses}
We assume that there exists a vector $\beta\in\R^T$ that can be used to assess the plausibility of the augmented PTA. For instance, in \eqref{main model new}, each $\beta_l$ corresponds to a "placebo" treatment effect in period $l\in\{1,...,T\}$. To keep the notation tractable, the dimension of $\beta$ corresponds to the number of available pre-treatment periods $T$. In practice, it is of course possible that the dimension of $\beta$ is smaller than $T$ (e.g.\ when only a subset of all pre-treatment period is used for a pre-test) or exceeds $T$ (e.g.\ when a conditional PTA is tested; see Section \ref{section heterogeneous}). Our tests can then be applied with minor adjustments.

Overall, we consider three distinct hypotheses to test for equivalence of pre-trends in treatment and control. We start with a discussion of the maximum placebo treatment effect. For $\beta\in\R^T$, level of significance $\alpha$ and the equivalence threshold $\delta>0$, we test
\begin{equation}
\mathrm{H}_0: \|\beta\|_{\infty}\geq\delta \qquad\mbox{ vs.} \qquad \mathrm{H}_1: \|\beta\|_{\infty}< \delta,
\label{equivalence hypothesis 1}
\end{equation}
where $\|\beta\|_{\infty}\defined \max_{l\in\{1,...,T\}}|\beta_l|$. 
Since we are now controlling the type I error, this implies that with confidence level of at least $1-\alpha$, $\delta$ is an upper bound for the maximum placebo treatment effect.

In many applications, pre- and post-treatment periods are pooled, for instance to increase statistical power. Similarly, it may be sensible in some applications to consider a pooled or average measure of the pre-treatment deviations from parallel trends. Thus, defining $\betabar\defined \frac{1}{T}\sum_{l=1}^{T}\beta_l$, one can find bounds on the average placebo effect by testing
\begin{equation}
\mathrm{H}_0: |\betabar|\geq \tau \qquad \mbox{vs.} \qquad \mathrm{H}_1: |\betabar|< \tau.
\label{equivalence hypothesis 2}
\end{equation}
One disadvantage of \eqref{equivalence hypothesis 2} is that there may be cancellation effects in situations where the components of $\beta$ are large in absolute terms but have opposing signs. Therefore, \eqref{equivalence hypothesis 2} should be used when differences in pre-trends can safely assumed to be of the same sign.  As pointed out in \cite{roth2020honest}, monotone violations of the PTA are frequently discussed in the applied literature. For instance, treatment effect estimates are often considered robust if potential violations of the PTA are of the opposing sign and can thus be ruled out as an explanation for the estimated effects. As an alternative to \eqref{equivalence hypothesis 2} that does not suffer from potential cancellation effects, we further consider the root mean square (RMS)  of $\beta$, i.e. 
$\beta_{RMS}\defined \|\beta\|/\sqrt{T}= \sqrt{\frac{1}{T}\sum_{l=1}^{T}\beta_l^2}$,
where $\|\cdot\|$ denotes the euclidean norm on $\R^{T}$. The RMS of $\beta$ can thus be interpreted as the euclidean distance between treatment and control in the pre-treatment periods relative to the distance in the base period scaled by the number of pre-treatment periods. The scaling is induced to ensure that this distance between treatment and control does not increase with the number of pre-treatment periods available. The hypothesis is then formulated  as
\begin{equation}
\mathrm{H}_0: \beta_{RMS}\geq\zeta \qquad \mbox{vs.} \qquad\mathrm{H}_1: \beta_{RMS}< \zeta ~,
\label{equivalence hypothesis RMS}
\end{equation}
which can equivalently be written as
\begin{equation}
\mathrm{H}_0: \beta_{RMS}^2 \geq \zeta^2 \qquad \mbox{vs.} \qquad\mathrm{H}_1: \beta_{RMS}^2 <  \zeta^2.
\label{equivalence hypothesis MS}
\end{equation}
In Section \ref{section Implementing equivalence tests} below we develop a test statistic for \eqref{equivalence hypothesis MS} and recover $\zeta$ as $\sqrt{\zeta^2}$.

\section{Theory and Implementation}
We begin this section by introducing the assumptions underlying our tests. We then proceed by discussing the implementation and the statistical properties of our tests.
\subsection{Assumptions}\label{assumptions}
For our statistical theory, we mainly need that a ``sequential'' version of asymptotic normality holds:
\begin{ass}\label{ass-sequential}
Fix $\varepsilon>0$.  For $\lambda\in[\varepsilon,1]$, let $\betahat (\lambda)$ an estimator of $\beta$ based on $\lfloor n  \lambda \rfloor$ individuals. Then
\begin{eqnarray} \label{hol25}
  \big \{ \sqrt{n} (\hat \beta (\lambda) - \beta ) \big \}_{\lambda \in [\varepsilon,1]}
    & \rightsquigarrow  & \big \{ \frac {1}{\lambda }  D  \vec{\mathbb{B} } (\lambda)\big \}_{\lambda \in [\varepsilon,1]},
\end{eqnarray}
where $ \vec{\mathbb{B} }$ is a $T$-dimensional vector of independent Brownian motions, $D$ is a $T \times T$-dimensional matrix of full rank,
 and "$\rightsquigarrow$" denotes weak convergence in the space $(\ell^\infty[\varepsilon,1])^{T}$ of all $T$-dimensional bounded functions on the interval $[\varepsilon,1]$.
\end{ass}
Notice that \eqref{hol25} comprises ``conventional" asymptotic normality of $\betahat$ when $\lambda=1$, i.e. 
\begin{equation}
\sqrt{n}(\betahat-\beta)\dto \mathrm{N}(0,\Sigma) \mbox{ as }n\to\infty,
\label{asymp norm}
\end{equation} where $\Sigma$ denotes a $T\times T$-dimensional covariance matrix. A further assumption required in some of our results is that this matrix  can be consistently estimated:
\begin{ass}\label{high-level 2}
An estimator $\hat{\Sigma}$ of $\Sigma$ is given such that
 $   \hat{\Sigma}\pto \Sigma \mbox{ as } n\to\infty$.
\end{ass}
The proofs of the validity of the first test for the hypothesis \eqref{equivalence hypothesis 1} and of the test for the hypothesis \eqref{equivalence hypothesis 2} do not require Assumption \ref{ass-sequential} but only the  weaker condition \eqref{asymp norm} together with Assumption \ref{high-level 2} (see the first part of Section \ref{max tests} 
and  Section \ref{average} below for the exact definition of these two tests). A second more powerful test for the hypothesis \eqref{equivalence hypothesis 1} is introduced in the second part of Section \ref{max tests}, and its validity will be established in the TWFE model \eqref{main model new}, under conditions that imply \eqref{asymp norm}.
In contrast, our test for the hypothesis \eqref{equivalence hypothesis RMS} requires \eqref{hol25}, but Assumption \ref{high-level 2} is not needed, as this test is based on ``self-normalization''. We emphasize that  without any additional assumptions the asymptotic normality   in  \eqref{asymp norm} does not imply  the process convergence in \eqref{hol25}. In fact, this a very delicate probabilistic question, which has,   to our best knowledge,  only been solved for sums of random variables. For example, we refer to \cite{kuelbs1973} for the independent case and to \cite{samur1987} for the dependent case (note that these authors consider Banach space valued random variables, which contains the case of finite dimensional vectors).
 
 Sequential asymptotic normality as in Assumption \ref{ass-sequential} can be shown to hold for stationary processes under various forms of dependence such as mixing or physical dependence (see \citealp{merlevede2006} and the references therein). Moreover, our tests can be flexibly adapted to different types of data (e.g.\ panel data or repeated cross-sections) and other features of the design such as clustering or staggered treatment assignment. As an illustration, in Appendix \ref{appendix illustration}, we verify that Assumption \ref{ass-sequential} holds under mild conditions in the canonical DiD model of Section \ref{section pre-testing in the canonical}.

Notice that our approach is a test of the PTA conditional on additional parametric restrictions (for instance, \eqref{no-anticipation new} and \eqref{Y0 model}, or parametric restriction on the effect of observed covariates). While arguably the vast majority of empirical studies are willing to impose similar assumptions, one might want to test the PTA without these restrictions. In this case, as an alternative to our method, one could consider tests based on conditional moment inequalities (see, among others, \citealp{andrews-soares-2010}, \citealp{romano-shaikh-wolf-2014} or \citealp[Section 4]{canay-shaikh-2017}).

\subsection{Implementing equivalence tests}\label{section Implementing equivalence tests}
Based on Assumptions \ref{ass-sequential} and \ref{high-level 2}, we now derive test statistics for our equivalence hypotheses. We further analyze the statistical properties of the resulting test procedures.
\subsubsection{Tests for the hypothesis \eqref{equivalence hypothesis 1}. }\label{max tests}
To describe the first test for the hypothesis \eqref{equivalence hypothesis 1} we initially consider the case  $T=1$ so that our objective is to test whether a single parameter $\beta_1$ exceeds a certain threshold.
As $\betahat_1$ is approximately distributed as $\mathrm{N}(\beta_1,\Sigma_{11}/n)$, the test statistic $|\betahat_1|$ approximately follows a folded normal distribution. We therefore propose to reject the null hypothesis in \eqref{equivalence hypothesis 1} whenever
\begin{align} \label{testt=2}
  |\betahat_1| <  \mathcal{Q}_{\mathrm{N}_F(\delta,\hat{\Sigma}_{11}/n)}(\alpha),
  \end{align}
   where $\mathcal{Q}_{\mathrm{N}_F(\delta,\sigma^{2})}(\alpha)$ denotes the $\alpha$ quantile of the folded normal distribution with mean $\delta$ and variance $\sigma^{2}$. It is shown in Appendix \ref{appendix mathematical proofs} that this test is consistent, has asymptotic level $\alpha$ and  is (asymptotically) uniformly most powerful for testing the hypothesis \eqref{equivalence hypothesis 1} in the case $T=1$. In particular this test is more powerful than the two-sided $t$-test (TOST), which could be developed following the arguments in \cite{hartman2018}.
  For $T>1$, we apply the idea of intersection-union (IU) tests outlined in \cite{berger1996bioequivalence} and reject
  the null hypothesis  in \eqref{equivalence hypothesis 1}, whenever
\begin{equation}
|\hat{\beta}_t| <  \mathcal{Q}_{\mathrm{N}_F(\delta,\hat{\Sigma}_{tt}/n)}(\alpha)~\forall t\in\{1,\dots,T\}.
\label{test statistic BOT}
\end{equation}
While this test is computationally attractive, a well-known disadvantage of testing procedures based on the IU principle is that they tend to be rather conservative \citep[see][among others]{berger1996bioequivalence}, which is confirmed by our simulation study (see 
Table \ref{rejection frequencies ar3 1}).

Specifically for the TWFE model \eqref{main model new}, it is possible to obtain a more powerful test for the hypothesis \eqref{equivalence hypothesis 1} as follows: In the first step, estimate $\beta$ in \eqref{main model new} to obtain the unconstrained TWFE estimator $\betahat_u$. In the second step, re-estimate \eqref{main model new} by minimizing the sum of squared residuals under the constraint $ \| \beta \|_\infty =
\max_{l=1,...,T}|\beta_l|
= \delta$  to obtain a constrained estimator, say $\betahat_c$. We then define new estimators of the parameters as 
\begin{equation} \label{MLcons}
{\hat{\betahat}_c}= \left\{
\begin{array} {ccc}
\hat \beta_u & \mbox{if} &    \|\betahat  \|_\infty 
\geq  \delta  \\
\hat \beta_c & \mbox{if} &  \|\betahat  \|_\infty 
<  \delta 
\end{array}  \right. \quad  
\end{equation}
and  
\begin{equation}\label{sigma-constrained}
\hat{\hat{\sigma}}_c=\frac{1}{(n-1)T}\sum_{i=1}^n\sum_{t=1}^{T+1}(\Yitdots-\Witdots'\hat{\betahat}_c)^2.
\end{equation}
Note that  the vector ${\hat{\hat{\beta}}_c}$ satisfies the null hypothesis in \eqref{equivalence hypothesis 1}. In the third step, for $b=1,...,B\in\N$, we generate bootstrap samples with $\uitdots^{(b)}\overset{\iid}{\sim}  \mathrm{N} (0,\hat{\hat{\sigma}}_c)$ and $\Yitdots^{(b)}=\Witdots'\hat{\hat \beta}_c+\uitdots^{(b)}$ for $i=1,...,n$ and $t=1,...,T+1$. 
For each bootstrap sample, estimate $\betahat^{(b)}$ and compute $\mathcal{Q}_\alpha^*$ as the empirical $\alpha$-quantile of the bootstrap sample $\{\max_{l=1,...,T}|\betahat_l^{(b)}| : b=1,...,B\}$. Finally, reject the null hypothesis $\Ho$ in \eqref{equivalence hypothesis 1} if 
\begin{align}
  \label{test statistic bootstrap}  
  \|\betahat\|_\infty <  \mathcal{Q}_\alpha^* ~.
\end{align}
Notice that the condition $ \| \beta \|_\infty
= \delta$ in the calculation of the constrained estimator $\hat \beta_c$ is made for technical reasons in the proof of this statement. Numerical results show that the test has a very similar behavior if  $\hat \beta_c$  is calculated under the condition that  $ \| \beta \|_\infty  \geq  \delta$. The following result shows that this test is consistent and  has asymptotic level
$\alpha$.

\begin{thm} \label{thmneu} 
If the errors $u_i=(u_{i,1},...,u_{i,T+1})'$ in model \eqref{main model new} satisfy $\E[u_i u_i'|G_i]=\sigma_u^2\mathrm{I}_{(T+1)\times (T+1)}$, then the test defined by \eqref{test statistic bootstrap} is consistent and has asymptotic level $\alpha $ for the hypothesis 
\eqref{equivalence hypothesis 1}. More precisely, 
\begin{itemize}
\item[(1)]
if  the null hypothesis in \eqref{equivalence hypothesis 1} is satisfied, then we have for any $\alpha \in (0, 0.5)$
\begin{equation}\label{level2.1}
\limsup_{n \rightarrow\infty}{\mathbb{P}_{\beta}}\big(
  \|\betahat\|_\infty <  \mathcal{Q}_\alpha^* 
\big)\leq\alpha.
\end{equation}
\item[(2)] if the null hypothesis in \eqref{equivalence hypothesis 1} is satisfied and the set 
\begin{align}
    \label{hd11}
\mathcal{E} = \{ \ell = 1, \ldots , T ~:~ |\beta_\ell| = \|\beta\|_\infty  \} 
\end{align}
consists of one point, then we have for any $\alpha \in (0, 0.5)$
\begin{equation*}
\lim_{n\rightarrow\infty}{\mathbb{P}_{\beta}}\big(
  \|\betahat\|_\infty <  \mathcal{Q}_\alpha^* \big) = \left\{
\begin{array} {ccc}
0 & \mbox{if} & \|\beta\|_\infty   > \delta  \\
\alpha & \mbox{if} & \|\beta\|_\infty  = \delta  .
\end{array} \right.
\end{equation*}
\item[(3)] if the alternative in \eqref{equivalence hypothesis 1}   is satisfied, then we have for any $\alpha \in (0, 0.5)$
\begin{equation*}
\lim_{n\rightarrow\infty} {\mathbb{P}_{\beta}}\big(
  \|\betahat\|_\infty <  \mathcal{Q}_\alpha^* \big) =1.
\end{equation*}
\end{itemize}
\end{thm}

\begin{rem} \label{det100}
{\rm  ~~
\begin{itemize}
    \item[(a)] 
It is well known
that  the bootstrap fails when the target parameter is non-differentiable, and so alternative
approaches are needed \citep[see, for example,][]{fang2019inference,HongLi2020}.
To explain why these observations are consistent with the statements in Theorem \ref{thmneu}, 
we briefly explain the main arguments for its proof here.  To be precise, note that  Theorem \ref{thmneu}
distinguishes  several cases under the null hypothesis.  First we consider the case (2), where  the set $
\mathcal{E} = \{ \ell = 1, \ldots , T ~:~ |\beta_\ell| = \|\beta\|_\infty  \} 
$
consists of one point. In this case, as    $\|\beta\|_\infty  \geq \delta >0$,    the mapping $\beta \to \|\beta\|_\infty$
is in fact Hadamard-differentiable, and one could alternatively  use Theorem 3.1 in \cite{fang2019inference} to prove that 
\begin{align}
\label{hd12}
\lim_{n\rightarrow\infty}{\mathbb{P}_{\beta}}\big(
  \|\betahat\|_\infty <  \mathcal{Q}_\alpha^* \big) = 
\alpha ,  
\end{align}
if  $\|\beta\|_\infty  = \delta >0 $, and  that the limit vanishes if  $\|\beta\|_\infty  > \delta$.
Next, we consider the case (1) of Theorem \ref{thmneu} and the situation where the  set $\mathcal{E} $ in \eqref{hd11} contains  more  than  $1$ element. In this case  
 the mapping $\beta \to \|\beta\|_\infty$ is not Hadamard-differentiable at this point. Consequently, by Theorem 3.1 in \cite{fang2019inference} a statement 
 of the from \eqref{hd12} cannot hold in general.
 However,  in  \eqref{level2.1} we neither claim the existence of the limit nor equality. In fact we can show by similar arguments as given for the derivation of equation (A.32) in \cite{detmolvolbre2015}  that 
$$
\sqrt{n} ( \|\betahat^{*}\|_\infty -  \|{\betahat}\|_\infty ) \leq \tilde Z_n^* + o_\mathbb{P} (1) ~,
$$
where the statistic $\tilde Z_n^*$ has, conditionally on the data, the same asymptotic distribution as the original statistic $\sqrt{n} ( \|{\betahat}\|_\infty -  \|{\beta}\|_\infty ) $. It turns out that  this argument is sufficient to prove \eqref{level2.1}, and the same argument can also  be used to show assertion (3) of Theorem \ref{thmneu}.
\item[(b)] 
We emphasize that Theorem \ref{thmneu} provides a pointwise result. We expect that uniform  results can be obtained as well. However, as indicated in part (a), even proving pointwise results requires some non-standard  techniques. This  is even more true if one would aim for uniform results, and we leave this problem for future research.

\item[(c)] 
The assumption  of  spherical model errors, i.e.\ $\E[u_i u_i'|G_i]=\sigma_u^2\mathrm{I}_{(T+1)\times (T+1)}$, is crucial for the validity of Theorem \ref{thmneu}. In fact, our simulation results in Section \ref{sec simulations} suggest that the test may not keep its nominal level under high levels of serial dependence in the error terms. As a further alternative, we therefore suggest replacing the bootstrap procedure described before \eqref{test statistic bootstrap} with the wild cluster bootstrap \citep{cameron2015}. To implement this, first compute the residuals $\hat{\ddot{u}}_{i,c}=\Yidots-\Widots'\hat{\betahat}_c$, where $\Yidots=(\ddot{Y}_{i,1},...,\ddot{Y}_{i,T+1})'$ and $\Widots=(\ddot{W}_{i,1},...,\ddot{W}_{i,T+1})$. Next, let $R_i$ denote a Rademacher random variable that takes the values $-1$ and $1$ with probability $0.5$ each. For $b\in\{1,...,B\}$, generate bootstrap samples 
$\Yidots^{(b)}=\Widots'\hat{\betahat}_c+ \hat{\ddot{u}}_{i}^*$, where $\hat{\ddot{u}}_{i}^*=R_i\times\hat{\ddot{u}}_{i,c}$. As before, compute $\betahat^{(b)}$ in each bootstrap sample, compute $\mathcal{Q}_\alpha^*$ and reject $\Ho$ in \eqref{equivalence hypothesis 1} if \eqref{test statistic bootstrap} holds. We demonstrate by means of a simulation study in Section \ref{sec simulations}  that this bootstrap test yields reasonable results for non-spherical errors, and we conjecture that the wild bootstrap provides a valid solution in general. A formal proof of this statement is beyond the scope of the present paper.
\end{itemize}}
\end{rem}

\begin{rem} \rm Theorem \ref{thmneu} shows the consistency of a specific estimator in model \eqref{main model new}. As pointed out by a referee, it is of interest to investigate if a similar approach is applicable for other estimators and models. We conjecture that this is possible, but proving the consistency of such an approach would be challenging, as the result would depend sensitively on the  given model and the estimator used.
\\
To be precise, let $\hat \beta $ denote an estimator of a parameter $\beta$, such that  $\sqrt{n}(\betahat-\beta)\dto \mathrm{N}(0,\Sigma) \mbox{ as }n\to\infty$, where $\Sigma$ denotes the  covariance matrix, and define the constrained  
estimate ${\hat{\betahat}_c}$ via \eqref{MLcons}.  Further let $\hat \Sigma$ denote a consistent estimate of the matrix $\Sigma$. We propose to generate $\betahat^{(b)}$ from a $\mathrm{N} ({\hat{\betahat}_c}, \hat \Sigma /n) $ distribution 
and compute $\mathcal{Q}_\alpha^{**}$ as the empirical $\alpha$-quantile of the bootstrap sample $\{\max_{l=1,...,T}|\betahat_l^{(b)}| : b=1,...,B\}$. The null hypothesis $\Ho$ in \eqref{equivalence hypothesis 1}  is rejected, if $ \|\betahat\|_\infty <  \mathcal{Q}_\alpha^{**}$. Note that implementation of this procedure neither requires a specific model nor a specific estimator.
In particular, we have implemented this alternative approach in  model \eqref{main model new} with 
$\hat\Sigma=\hat{\hat{\sigma}}_c^2 (\ddot{W}'\ddot{W})^{-1}$ for spherical errors and with $\hat\Sigma=\frac{n}{n-T}(\ddot{W}'\ddot{W})^{-1}(\sum_{i=1}^n \ddot{W}_i\hat{\ddot{u}}_{i,c}\hat{\ddot{u}}_{i,c}'\ddot{W}_i')(\ddot{W}'\ddot{W})^{-1}$ for errors that are clustered on the individual level, where $\ddot{W}=(\ddot{W}_1,...,\ddot{W}_n)'$ and $\hat{\hat{\sigma}}_c^2$ and $\hat{\ddot{u}}_{i,c}$ are defined in \eqref{sigma-constrained} and Remark \ref{det100}, respectively. In a simulation study, we obtained very similar results as for the bootstrap tests considered in Theorem \ref{thmneu} for the spherical case and Remark \ref{det100} (c)  for the clustered case (these results are not displayed in Section \ref{sec simulations} for the sake of brevity).

\end{rem}
\subsubsection{A test for the hypothesis \eqref{equivalence hypothesis 2}. }\label{average}
 For some fixed $\tau>0$,  a test can be constructed
 by first computing the statistic
\[
\bar{\betahat}\defined\frac{1}{T}\sum_{t=1}^{T}\betahat_t = \mathds{1}'\betahat/T,
\]
where $\mathds{1} = (1,\ldots , 1 )' \in \R^{T}$. Note that it follows from Assumptions \ref{ass-sequential} and \ref{high-level 2} that $$\sqrt{n} \mathds{1}'({\hat \beta} - \beta) \to \mathrm{N}(0, \mathds{1}' \Sigma \mathds{1})).$$
Consequently, based on the discussion in the first part of \ref{max tests}, we propose to reject the null hypothesis in
\eqref{equivalence hypothesis 2},  whenever
\begin{equation}
|\bar{\betahat}|  <  \mathcal{Q}_{\mathrm{N}_F(\tau, \hat{\sigma}^{2})}(\alpha),~
\label{test statistic mean beta}
\end{equation}
where $\hat{\sigma}^{2}=\mathds{1}'{\hat{\Sigma}} \mathds{1}$/n, and $\hat{\Sigma}$ is a consistent estimator of $\Sigma$ which is known by Assumption \ref{high-level 2}.

\subsubsection{A pivotal test for the hypothesis \eqref{equivalence hypothesis MS}. }
For $\lambda\in [\varepsilon,1]$, let $\betahat_{RMS}(\lambda)$ denote the RMS of $\betahat$ based on a random subsample of size $\lfloor \lambda n\rfloor$ of the original sample of $n$ individuals. In order to construct a pivot test for the hypothesis \eqref{equivalence hypothesis MS}, define 
\begin{equation}\label{mt}
\hat M_n \defined  \frac{\hat{\beta}_{RMS}^2(1)-  \beta_{RMS}^2}{\hat{V}_n},
\end{equation}
where
\begin{equation}\label{definition VNsq}
\hat{V}_n=  \Big (\int_\varepsilon^1 (\hat{\beta}_{RMS}^2(\lambda)-\hat{\beta}_{RMS}^2(1))^2\nu(d\,\lambda) \Big )^{1/2}
\end{equation}
and $\nu$ denotes a measure on the interval $[\varepsilon,1]$.  The following result is proved in the Appendix. 

\begin{thm} \label{thm0}
 If  Assumption  \ref{ass-sequential} is  satisfied and $\beta\not =0$, then the statistic $\hat M_n$ defined  in \eqref{mt} 
 converges weakly with a non-degenerate limit distribution, that is
\begin{equation}
\hat M_n \dto \mathbb{W}:= \frac{ \mathbb{B}(1)}{\big(\int_\varepsilon^1(\mathbb{B}(\lambda)/\lambda-\mathbb{B}(1))^2 \nu(d \lambda) \big)^{1/2}
},
\label{asymptotic distribution TN}
\end{equation}
where $\{ \mathbb{B}(\lambda)\}_{\lambda \in [\varepsilon,1] }$ is a Brownian motion on the interval $ [\varepsilon,1]$.  Moreover, if $\beta =0$, then 
\begin{align}
   \label{hol31}
\hat M_n \dto \frac{ \mathbb{Z}^2(1)}{\big(\int_\varepsilon^1(\mathbb{Z}^2(\lambda) -\mathbb{Z}^2(1))^2 \nu(d \lambda) \big)^{1/2}}~,
\end{align}
where $\mathbb{Z}^2(\lambda) = \frac{1}{\lambda^2}  \vec{\mathbb{B}}'(\lambda) D' D  \vec{\mathbb{B}}(\lambda)
$, $\vec{\mathbb{B} }$  is a   $T$-dimensional vector of independent Brownian motions and $D$ is the matrix in \eqref{hol25}.
\end{thm}
\noindent
It follows from the proof of Theorem \ref{thm0} that the statistic $\hat{\beta}_{RMS}^2$ is a consistent estimator of $\beta_{RMS}^2$. Therefore, we propose to reject the null hypothesis $\mathrm{H}_0$ in \eqref{equivalence hypothesis MS} (and consequently $\mathrm{H}_0$ in \eqref{equivalence hypothesis RMS}),  whenever
\begin{equation}
\hat{\beta}_{RMS}^2 
<\zeta^2 +\mathcal{Q}_{\mathbb{W}}(\alpha) \hat{V}_n,
\label{test statistic rms}
\end{equation}
where $\mathcal{Q}_{\mathbb{W}}(\alpha)$ is the $\alpha$-quantile of the limiting distribution of the random variable $\mathbb{W}$ on the right-hand side of \eqref{asymptotic distribution TN}. Note that these quantiles  can be easily obtained by simulation because the distribution of $\mathbb{W}$ is completely known.  For instance, $\mathcal{Q}_{\mathbb{W}}(0.05)\approx -2.1$. The following result shows that this decision rule defines a valid test for the hypothesis \eqref{equivalence hypothesis MS}.
\begin{thm} \label{thm1}
If Assumption \ref{ass-sequential} is satisfied, then the test defined by \eqref{test statistic rms} is a consistent asymptotic level $\alpha$-test for the hypothesis \eqref{equivalence hypothesis MS}, that is
\[
\lim_{n \to \infty} {\mathbb{P}_{\beta }} \Big(\hat{\beta}_{RMS}^2 < \zeta^2 +\mathcal{Q}_{\mathbb{W}}(\alpha) \hat{V}_n\Big)
= \left\{
 \begin{array}{ccc}
   0, & \mbox { if } & \beta_{RMS}^2 > \zeta^2 \\
   \alpha, & \mbox { if } & \beta_{RMS}^2 = \zeta^2 \\
   1, & \mbox { if } & \beta_{RMS}^2 < \zeta^2
 \end{array} \right .
\]
\end{thm}

\begin{rem} ~~~ \\ 
{\rm 
(a)
 Notice that in practice one chooses $\nu$ as a discrete distribution which makes the evaluation of the integrals in \eqref{definition VNsq} and in the denominator of the
 random variable $ \mathbb{W}$
  very easy. For example,
   if $\nu$ denotes the uniform distribution on $\{\frac{1}{5},\frac{2}{5},\frac{3}{5},\frac{4}{5}\}$, then the statistics $\hat{V}_n$ in  \eqref{definition VNsq} simplifies to 
\[
\Bigl(\frac{1}{4}\sum_{k=1}^4 (\betahat_{RMS}^2  (\tfrac{k}{5})-\betahat_{RMS}^2  (1))^2\Bigr)^{1/2}~.
\]
This measure is also used in the  simulation study in Section \ref{sec simulations}, where we analyze the finite sample properties of the different procedures. In practice, it is thus not necessary to explicitly choose $\varepsilon$.
\\
(b)  It follows from the proof of Theorem \ref{thm0} that an asymptotic $(1-\alpha)$-confidence interval for the parameter $\beta_{RMS}^2 >0 $ is given by 
$$
\Big [ \betahat_{RMS}^2  +\mathcal{Q}_{\mathbb{W}}(\alpha/2) \hat{V}_n,   
~ \betahat_{RMS}^2  +  \mathcal{Q}_{\mathbb{W}}(1-\alpha/2) \hat{V}_n
\Big ]. 
$$
(c) Theorem \ref{thm1} can be extended to get uniform results. More precisely, define for a small  positive constant $c$ the sets
\begin{align*}
{\cal H} &= \Big \{ \beta  ~\Big |~
   \Delta  (\beta ) > c~,~ 
\betahat_{RMS}^2  \geq \zeta^2  \Big \}~, \\
    {\cal A} (x)  &= \Big \{ \beta  ~\Big |~
   \Delta  (\beta ) > c~,~ 
 \betahat_{RMS}^2 <  \zeta^2  - x / \sqrt{n}  \Big \} ~, 
\end{align*}
corresponding to the null hypothesis and the alternative, respectively, where $ \Delta  (\beta ) $ is defined in equation
\eqref{hol21} in the Appendix. Then 
$$
 \limsup_{n \to \infty} \sup_{\beta   \in \mathcal{H}}
 \mathbb{P}_{\beta } \Big(
\betahat_{RMS}^2 <   \zeta^2 + \mathcal{Q}_{\mathbb{W}}(\alpha) \hat{V}_n 
 \Big)=\alpha ~.
 $$
 Furthermore there exists a non-decreasing function $f: \mathbb{R}_{>0} \to \mathbb{R}_{>0}$, with $f(x)> \alpha$ for all $x >0$ and $\lim_{x \to \infty}f(x)=1$, such that
 $$
 \liminf_{n \to \infty} \inf
 _{\beta   \in \mathcal{A} (x)  } \mathbb{P}_{\beta } \Big(
 \betahat_{RMS}^2 <  \zeta^2  + \mathcal{Q}_{\mathbb{W}}(\alpha) \hat{V}_n
 \Big)
 =f(x)~.
 $$
 The details are omitted for the sake of brevity. 
 }
\end{rem}

\section{Equivalence testing with heterogeneous treatment effects and staggered adoption}\label{section heterogeneous}
The use of the simple TWFE model has recently experienced substantial criticism in the presence of multiple groups, heterogeneous treatment effects and differences in treatment timing. In this situation, the TWFE estimator often does not correspond to a reasonable estimate of the ATT, and alternative estimators have been proposed by several authors (see, for instance, \citealp{goodman2018difference}, \citealp{callaway2020difference},  \citealp{abraham2018estimating}, \citealp{borusyak} or \citealp{chaisemartin2020}. Excellent reviews of this fast-growing literature are provided by \citealp{roth-survey} and \citealp{chaisemartinDiD}.). Specifically, \citet{wooldridge2021two} shows that the deficiency of the TWFE estimator can be regarded as a model misspecification problem. He then proposes model adjustments that allow for treatment effect heterogeneity due to differences in treatment timing and observed characteristics (which are assumed to be unaffected by the treatment). While we conjecture that our equivalence tests can be used with most (if not all) of the mentioned estimators, we focus on the regression-based approach of \citet{wooldridge2021two} in the following, as it is straightforward to show that our assumptions hold with minor adjustments to the arguments in Appendix \ref{appendix illustration}.
We now consider the case of the staggered adoption (i.e.\ the initial treatment period varies across groups) of an absorbing treatment (i.e.\ the treatment status does not change after the initial treatment) in the presence of a never-treated group. Following \citet[Section~6]{wooldridge2021two}, we assume that the time since the initial treatment adoption produces different levels of exposure to the treatment, resulting in treatment effect heterogeneity across time. As in Section \ref{section pre-testing in the canonical}, we consider a balanced panel of $n$ individuals that are observed in $T+1$ pre-treatment periods. In periods $T+2,...,\overline{T}$, a subset of individuals adopts treatment, leading to ``treatment cohorts''. As before, period $T+1$ is used as the base period. To define a treatment cohort dummy, let $G_i^r=1$ if individual $i$ has first adopted treatment  in period $r\in\calR\defined\{T+2,...,\overline{T},\infty\}$ and zero otherwise, where $G_i^\infty$ is a dummy indicating that individual $i$ is a member of the never treated group. 
The potential outcome of unit $i$ in treatment cohort $r\in\calR$ observed in time period $t\in\{1,...,\overline{T}\}$ is denoted by $Y_{i,t}(r)$, where the ``baseline'' potential outcome in period $t$ if unit $i$ is untreated is given by $Y_{i,t}(\infty)$. The objects of interest are the cohort and time-specific treatment effects that may depend on a vector of observed and time-invariant covariates $X_i$, i.e.
\begin{equation}
\pi_{r,t}(X_i)\defined \E[Y_{i,t}(r)-Y_{i,t}(\infty)|G_i^{r}=1,X_i].
\label{ATT staggered}
\end{equation}
In many applications, it is assumed that the treatment effect is a linear function of the observed covariates (which may contain polynomials of $X_i$), so that
\begin{equation*}
\pi_{r,t}(X_i)=\pi_{r,t} + \rho_{r,t}'\dot{X_i},
\end{equation*}
where $\dot{X}_i=X_i-\E[X_i|G_i^r=1]$ is the covariate centered around the cohort mean so that $\pi_{r,t}$ is the treatment effect averaged across the distribution of $X_i$ conditional on $G_i^r=1$.  Adapting \eqref{no-anticipation new}, we assume 
\begin{equation*}
\E[Y_{i,t}|\Gi,X_i]=\E[Y_{i,t}(\infty)|\Gi,X_i]+\sum_{r=T+2}^{\overline{T}}\sum_{s=r}^{\overline{T}}\pi_{r,s}(X_i)G_i^{r}D_{s}(t),
\end{equation*}
where $\Gi=(G_i^{T+2},...,G_i^{\overline{T}})$, so that deviation from the designated treatment path (e.g.\ through anticipation) are ruled out. Further extending \eqref{Y0 model},
\begin{equation*}
\E[Y_{i,t}(\infty)|\Gi,X_i]=\alpha_i+\lambda_t 
+\kappa'X_i
+\sum_{r=T+2}^{\overline{T}} \zeta_r'X_iG_i^r
+ \sum_{\substack{s=1\\s\neq T+1}}^{\overline{T}}\iota_s'X_iD_s(t)
+\sum_{\substack{r=1\\r\neq T+1}}^{\overline{T}}\sum_{\substack{s=1\\s\neq T+1}}^{\overline{T}}\gamma_{r,s}G_i^rD_s(t),
\end{equation*}
where $\gamma_{r,s}$ is a non-stochastic constant that differs across cohorts and time. Notice that the covariates are again assumed to enter the baseline outcome linearly. As in Section \ref{section pre-testing in the canonical}, the cohort and time-specific ATTs cannot be identified without further restrictions. In order to ensure that $\gamma_{r,s}=0$, we adapt
\eqref{statement parallel trends new} to impose a conditional staggered parallel trends assumption (CSPTA), using the never-treated group as the control:
\begin{equation*}
\E[Y_{i,t}(\infty)-Y_{i,T+1}(\infty)|G_i^r=1,X_i]=\E[Y_{i,t}(\infty)-Y_{i,T+1}(\infty)|G_i^\infty=1,X_i],~t,r=1,...,\overline{T}.
\end{equation*}
Combining the above assumptions, we obtain
\begin{align}\label{staggered model}
&\E[Y_{i,t}|\Gi,X_i]=\alpha_i+\lambda_t+\kappa'X_i\sum_{r=T+2}^{\overline{T}}\zeta_r'X_i G_i^r 
+\sum_{\substack{s=1\\s\neq T+1}}^{\overline{T}} \iota_s'X_i D_{s}(t)
\notag\\&+ \sum_{r=T+2}^{\overline{T}}\sum_{s=r}^{\overline{T}}(\pi_{r,s}G_i^r D_{s}(t) + \rho_{r,s}'\dot{X}_i G_i^rD_s(t))
+\sum_{m=T+2}^{\overline{T}}\sum_{\substack{k=1\\k\neq T+1}}^{m-1}(\tilde{\pi}_{m,k}G_i^m D_{k}(t) + \tilde{\rho}_{m,k}'\dot{X}_i G_i^mD_k(t)).
\end{align}
Simple algebra sows that the placebo conditional cohort and time specific treatment effects $\tilde{\pi}_{m,k}+\tilde{\rho}_{m,k}'\dot{X}_i$ are identified by taking differences-in-differences, i.e.\ by comparing the evolution in the average outcome between periods $k$ and the base $T+1$ between treatment cohort $m$ and the never treated group. Given \eqref{ATT staggered}--\eqref{Y0 model}, the CSPTA implies that the placebo treatments are zero. We therefore avoid any ``contamination'' by treatment effects at time $m'>m$, which, as noted by \cite{abraham2018estimating}, can lead to a rejection of the CSPTA in the pre-treatment periods even in cases where it actually holds.
Since the assumptions in Section \ref{assumptions} can easily be shown to hold under mild conditions by adapting the arguments in Section \ref{appendix illustration}, we can directly apply our equivalence tests to the placebo treatment effects. Moreover, the model in \eqref{staggered model} can be flexibly adjusted to the problem at hand. For instance, one may be willing to exclude a subset of the placebo treatment effects from the model in order to allow for some pooling across cohorts and time. As noted by \cite{wooldridge2021two}, in this case, the pooled OLS estimator of $\pi_{r,s}$ is an averaged ``rolling DiD'' where, on top of the never-treated group and the base period, any cohort and period that corresponds to an omitted placebo treatment effect is used as a control. In this case, the CSPTA needs to be adjusted accordingly (e.g.\ as in \citealp{roth-survey}), as parallel trends need to be plausible between multiple groups and periods. Finally, notice that in practice $\E[X_i|G_i^r=1]$ needs to be replaced by the sample average of $X_i$ in cohort $r$. As suggested in \cite{wooldridge2021two}, one should adjust the standard errors to account for the additional sampling variation.

As a further note of caution, practitioners should be aware that our tests only consider an implication of the CSPTA, as we are relying on assumptions that impose a particular form of treatment effect heterogeneity (e.g.\ linear dependence on observed covariates). In fact, we are thus testing a joint
test of a parametric restriction on treatment effect heterogeneity and conditional parallel trends. While parametric restrictions are popular in the applied literature, one might still be concerned about their validity when testing for parallel trends. As an alternative, one may then consider tests based on conditional moment inequalities (see, for instance, \citealp{andrews-shi}).

\section{Simulations}\label{sec simulations}
In order to investigate the small sample properties of our tests, we conduct a simulation study in \textbf{R}. For that, we create a panel data set with $T\in\{1,4,8,12\}$ and $n\in\{100,1000\}$.  For $i=1,...,n$ and $t=1,...,T+2$, we generate the data from 
\begin{equation}\label{simulation model}
    Y_{i,t}=\alpha_i+\lambda_t + \sum_{l=1}^{T}\beta_lG_i D_{l}(t)+\pi_{ATT}G_i D_{T+2}(t) +u_{i,t},
\end{equation} 
with $\alpha_i$ and $\lambda_t$ standard normal, $\Pr(G_i=1)=\frac{1}{2}$ and $\pi_{ATT}=0$.  

In an initial step, we investigate the level of the proposed tests. To do so, we set the level of significance  to $\alpha=5\%$ and the equivalence threshold for all hypotheses to $1$. We then choose the parameters $\beta_l$ in the pre-treatment periods such that we are on the ''boundary'' of the hypotheses, that is $\beta_1=1$ and $\beta_l=0$ for $l>1$ or $\beta_l=1$ for all $l=1,...,T$. Moreover, we also investigate the power of the test procedures by choosing  $\beta_l\in\{0.8,0.9\}$ for all $l=1,...,T$. The bootstrap based tests for \eqref{equivalence hypothesis 1} are computed using $1000$ bootstrap draws. Finally, the generated error terms are both serially correlated and heteroskedastic, as they are drawn from a stationary AR(3) process with autoregressive parameters $(\phi_1,\phi_2,\phi_3)=(0.5,0.3,0.1)$ and with standard deviation $1+G_i$. Consequently, the tests that require an estimate of $\Sigma$ are based on standard errors that are clustered on the individual level. The results for all tests based on $20000$ simulations are presented in Tables  \ref{rejection frequencies ar3 1}--\ref{power all ar3}.

In the following scenarios, we choose the level of significance $\alpha=5\%$ and compute $\delta^*_{IU}$, $\delta^*_{Boot}$ and $\delta^*_{c.Boot}$ as the smallest equivalence thresholds for the IU, the bootstrap and the cluster bootstrap tests such that the null hypothesis in \eqref{equivalence hypothesis 1} can still be rejected (i.e.\ for which equivalence of pre-trends can be concluded). Similarly, we compute the smallest equivalence thresholds $\tau^*$ and $\zeta^*$ for the corresponding null hypotheses in  \eqref{equivalence hypothesis 2}  and \eqref{equivalence hypothesis RMS} using the tests in \eqref{test statistic mean beta} and \eqref{test statistic rms}, respectively. The reported numbers correspond to the average over $M=2500$ simulations and can be used to assess at what value of the equivalence threshold a particular test can be expected to reject the null hypothesis. Finally, we report the usual $95\%$ confidence interval $\mathrm{CI}^{\hat{\pi}_{ATT}}$ and the number of simulations in which each $\beta_l$ for $l=1,...,T$ was found to be statistically insignificant. We then investigate how violations of the PTA affect the chance of falsely detecting a treatment effect and how these violations affect the smallest equivalence thresholds for which equivalence can be concluded. For simplicity, the model errors are drawn independently from a standard normal distribution. 

Table \ref{simulation pta homoskedastic} shows the results under the PTA. We further simulate scenarios in which the PTA is violated due to the presence of unobserved effects that affect the treatment group but not the control group. First, we simulate a pre-program shock, also known as ``Ashenfelter's dip'' by replacing $\uit$ in \eqref{simulation model} by $\tilde{u}_{i,t}=\uit+ G_i D_{T+1}(l) V_i$, where $V_i\sim \mathrm{N} 
(\mu,1)$ with mean $\mu\in\{\frac{1}{4},\frac{1}{2}\}$. This implies that $\E[\hat{\pi}_{ATT}]=\E[\betahat_l]=-\mu$ for $l=1,...,T$. The results are presented in Tables \ref{simulation ashenfelter 025} and \ref{simulation ashenfelter 05}. In our second scenario, $\tilde{u}_{i,t}=\psi\times t\times G_i$ with $\psi=0.025$, modeling an unobserved linear difference in time trends, starting in $t=1$. Consequently, $\E[\betahat_l]=\psi(l-T-1)$ and $\E[\hat{\pi}_{ATT}]=\psi$. The results are presented in Table \ref{time trend 0025 new homoskedastic}.
\subsection{Simulation results -- Discussion}
Table \ref{rejection frequencies ar3 1} shows that the test in \eqref{test statistic mean beta} approximately keeps the desired level for every $T$ even in small samples. The test in \eqref{test statistic rms}  appears to be slightly over-rejecting when $n =100$ but keeps its nominal level in larger samples.  Notice that in Table \ref{rejection frequencies ar3 2} the tests in \eqref{test statistic mean beta} and \eqref{test statistic rms} rightfully reject the null hypothesis in an increasing number of cases as $n $ and $T$ increase as $\betabar$ and $\beta_{RMS}$ are further away from the boundary of the null, resulting in an increase in statistical power. Regarding the tests for \eqref{equivalence hypothesis 1} Tables \ref{power all ar3} and \ref{power single ar3} illustrate that the IU and cluster bootstrap based tests maintain their nominal level for $T=2$. When only one parameter is at the boundary of the null hypothesis, both tests also perform well in the sense that the empirical rejection frequency is close to the nominal level for sufficiently large $n$. In comparison, the test in \eqref{test statistic bootstrap} over-rejects even for large $n$, showing that it is not robust to high levels of serial correlation. If $\beta_l=1$ for all pre-treatment periods, all three tests become conservative for larger values of $T$. This phenomenon appears to be much more pronounced for the test based on  the IU principle, for which it is well-documented \citep{berger1996bioequivalence}. For instance, the empirical level of the IU based test is more than 7 times smaller than the corresponding level of the bootstrap based tests for $T=12$ (see Table \ref{rejection frequencies ar3 1}). As shown in Table \ref{power all ar3}, this has important consequences for the power of both tests as the cluster bootstrap based test procedure outperforms the IU based test for $T>2$. On the other hand, the IU based test may still be attractive for practical applications as it is numerically much less demanding. As compared to the tests for \eqref{equivalence hypothesis 1}, the power of our test in \eqref{test statistic rms} is substantially larger, only surpassed by the power of the test in \eqref{test statistic mean beta}. All tests have in common that the power decreases with $T$. This is true even for the test in \eqref{test statistic mean beta}, which is the (asymptotically) uniformly most powerful test for \eqref{equivalence hypothesis 2} for any $T$. Thus, concluding equivalence of pre-trends becomes more demanding with an increase in the number of pre-treatment periods. This makes intuitive sense in the DiD setup, where equivalence of pre-trends in a larger number of periods is often regarded as stronger evidence for the plausibility of the PTA. 

For $T=1$, we find that $\zeta^*>\delta^*_{Boot}\approx \delta^*_{c.Boot}>\delta^*_{IU}=\tau^*$. This is not surprising, since the IU based test is asymptotically uniformly most powerful for the hypothesis \eqref{equivalence hypothesis 1}, where \eqref{equivalence hypothesis 1} coincides with \eqref{equivalence hypothesis 2} and \eqref{equivalence hypothesis RMS} when a single pre-treatment parameter is tested. For $T>1$, we roughly observe that $\tau^*<\zeta^*\leq\delta^*_{Boot}\leq \delta^*_{c.Boot} <\delta^*_{IU}$. This relationship between the smallest equivalence thresholds of the maximum, average and RMS tests is expected, as $|\betabar|\leq \beta_{RMS}\leq \|\beta \|_{\infty}$. The better performance of the bootstrap based tests for \eqref{equivalence hypothesis 1} as compared to the IU based test may attributed to their higher power as evidenced in Tables \ref{power all ar3} and \ref{power single ar3}. We also observe that $\delta^*_{Boot}$ performs only slightly better than $\delta^*_{c.Boot}$ under the PTA when errors are spherical. Under violations of the PTA, $\delta^*_{Boot}<\delta^*_{c.Boot}$ for $T>1$. However, as $\tilde{u}_{i,t}$ becomes non-spherical, only the cluster bootstrap based test maintains its nominal level (see Table \ref{rejection frequencies ar3 2}). We therefore recommend to use the cluster-robust version of the bootstrap based test in practice. Further notice that even when the PTA holds, the practice of rejecting the DiD framework when $\betahat_l$ is statistically insignificant for at least one $l\in\{1,...,T\}$ is clearly inefficient as is shown by the first row of Table \ref{simulation pta homoskedastic}, as an increase in available pre-treatment periods increases the chance of incorrectly rejecting the DiD framework under the PTA. A similar observation can be made in the presence of a linear time trend as shown in Table \ref{time trend 0025 new homoskedastic}. Here, even when the empirical coverage rate of the usual confidence interval is only slightly lower than the nominal level, the DiD framework is rejected in a large number of cases.

When the PTA is violated due to a small temporary shock as in Table \ref{simulation ashenfelter 025}, the usual practice of adopting the PTA when no significant differences in pre-trends could be found can lead to a false discovery of a non-zero treatment effect in a substantial number of cases, in particular when the sample size is small. If the temporal shock is larger as in Table \ref{simulation ashenfelter 05}, a non-existing treatment effect will be found to be significantly different from zero in almost all cases. All our test procedures require an unrealistically large equivalence threshold in order to be able to conclude equivalence of pre-trends. In particular, any equivalence threshold for which equivalence could be concluded would have to be larger than the estimated treatment effect, therefore casting serious doubt on the validity of the latter. 
Similarly, when the PTA is violated due to a linear difference in trends (see Table \ref{time trend 0025 new homoskedastic}), the equivalence thresholds would have to be chosen larger than the estimated treatment effect, thus suggesting that the estimated ATT may contain bias due to insufficient support for the PTA. Moreover, our methodology can be useful in identifying the presence of a linear time trend, as $\tau^*$ and $\zeta^*$ tend do remain stable in $T$ under the PTA or when the violation of the PTA is only temporary, whereas under the presence of a linear trend, they increase with $T$.

\section{Empirical illustration}
We illustrate our approach by re-considering the influential Difference-in-Differences analysis in \cite{ditellacrime2004}. They use a shock to the allocation of police forces as a consequence of a terrorist attack on a Jewish institution as a natural experiment to study the the effect of police on crime. The original authors conduct the usual pre-test in \eqref{individual H0} and find no evidence for violations of the PTA. However, \citet{donohue2013police} point out several shortcomings of the original paper (e.g.\ spillover effects from the treated to the untreated group). In particular, they find that  the PTA is not plausible if the pre-treatment data is inspected on a more granular level, thus casting doubt on the validity of the estimated treatment effects. While the traditional test failed to detect evidence \textit{against} the PTA, we will apply our test procedures to analyze how much evidence \textit{in favor} of the PTA can be extracted from the original specification in \cite{ditellacrime2004}.

The data consists of monthly averages of the number of car thefts between April and December 1994 in each out of 876 Buenos Aires city blocks out of which 37 blocks received additional protection after the attack. The main specification in \cite{ditellacrime2004} is given by $Y_{it}=\alpha_i+\lambda_t+\beta D_{it}$, where $Y_{it}$ denotes the number of car thefts in block $i$ and month $t$ and $D_{it}$ is a dummy variable taking the value 1 if block $i$ is treated in period $t$. Finally, $\alpha_i$ and $\lambda_t$ are block- and time-specific fixed effects. By using this specification, the pre- and post-treatment periods are pooled together so that the estimated treatment effect compares the post-treatment difference in car thefts between treated and non-treated blocks to the corresponding pre-treatment difference. To analyze group mean differences in the pre-treatment periods, we fit \eqref{main model new} to subsets of the data that include one, two or three pre-treatment periods, corresponding to June, May and June and April--June. As in the original paper, we include block- and time-specific effects and cluster on the block level. We find that using heteroskedasticity-robust standard errors instead of clustering has no substantial effect. Finally, we compute $\delta^*_{IU}$, $\delta_{Boot}^*$, $\delta_{c.Boot}^*$, $\tau^*$ and $\zeta^*$. The results are summarized in Table \ref{empirical results} below.
\begin{table}[h]
\center
\begin{tabular}{l||*{3}{c}}\toprule
\backslashbox{Estimates}{periods}
&\makebox[5em]{June}&\makebox[5em]{May\&June}&\makebox[5em]{April--June}
\\\hline\hline
$\delta_{IU}^*$&0.147&0.147&0.147\\
$\delta_{Boot}^*$ &0.156&0.156&0.15\\
$\delta_{c.Boot}^*$ &0.152&0.141&0.137\\
$\tau^*$ &0.147&0.086&0.093\\
$\zeta^*$ &0.152&0.12&0.129\\
\bottomrule
\end{tabular}
\caption{Smallest equivalence thresholds such that the null hypotheses in \eqref{equivalence hypothesis 1}, \eqref{equivalence hypothesis 2} and \eqref{equivalence hypothesis RMS} can be rejected for varying numbers of pre-treatment periods.}\label{empirical results}
\end{table} 
Notice that for all tests the smallest equivalence threshold that still allows us to conclude equivalence of pre-trends are the largest  when only the pre-treatment period June is used. This hints towards a temporary shock to treatment or control in June which may bias the pooled estimates in Table 3 of \cite{ditellacrime2004}. The latter are significant and range between $-0.058$ and $-0.081$. One important outcome of our equivalence test based analysis is that, even without the granular data inspection of \cite{donohue2013police}, the equivalence thresholds have to be chosen unrealistically large in order to conclude equivalence of pre-trends. In fact, the smallest equivalence thresholds for which the null hypotheses can be rejected are larger than the estimated effect size of police on crime. Therefore, it is questionable whether there is any effect at all, since the estimated effect may be an artifact of the violated PTA only.

\section{Conclusion}
We have derived several tests that allow researchers to provide statistical evidence in support of the parallel trends assumption in difference-in-differences estimation by testing for negligible differences in pre-trends in treatment and control. The tests can easily be implemented in popular models such as the two-way fixed effects model, and they can be flexibly adjusted to accommodate more complex settings. Our simulation analysis yields support for our theoretical results as our tests maintain their nominal level and exhibit high statistical power in sufficiently large samples. Finally, we apply our methodology to the data provided by \cite{ditellacrime2004}. Even without a granular inspection of the data as in \cite{donohue2013police}, our methodology casts doubt on the estimated effects, as they may simply be the result of trend differences that could not be detected using traditional pre-tests.

\section*{Acknowledgements}
We thank the editor Ivan Canay, an associate editor, and two anonymous referees for comments that greatly improved this paper. We further thank participants of the International Panel Data Conference 2023, the European Summer Meeting of the Econometric Society 2023 and the European Winter Meeting of the Econometric Society 2023.  

\section*{Conflict of interest statement}
The authors report there are no competing interests to declare.
\clearpage
\bibliographystyle{agsm}

\bibliography{mybib}

\newpage
\newcommand{\calS}{\mathcal{S}}
\appendix
\section{Assumption \ref{ass-sequential} in the canonical DiD model of Section \ref{section pre-testing in the canonical}}\label{appendix illustration}
For $\lambda\in [\varepsilon,1]$, compute the double-demeaned variables based on a random subsample $\calS_{n,\lambda}$ of $\lfloor n \lambda \rfloor$ individuals, each observed in periods $1,...,T+1$.  For instance, we define the double-demeaned data as $\ddot{Y}_{i,t}(\lambda)=Y_{i,t}-\Ybaridot-\Ybardott(\lambda)+\Ybardotdot(\lambda)$, where
\[
\Ybardott(\lambda)=\frac{1}{\lfloor n \lambda \rfloor}\sum_{j\in\calS_{n,\lambda}}Y_{j,t} ~\mbox{ and }~~
\Ybardotdot(\lambda)=\frac{1}{\lfloor n \lambda \rfloor (T+1)}\sum_{j\in\calS_{n,\lambda}}\sum_{s=1}^{T+1}Y_{j,s},
\]
and other variables such as $\ddot{W}_{i,t} (\lambda)$ and $\ddot{u}_{i,t} (\lambda) $ are defined analogously. Collecting the individual time series, 
\begin{equation}\label{main demeaned model old}
\Yidots(\lambda)=\Widots(\lambda)'\beta+\uidots(\lambda),
\end{equation}
where $\Yidots(\lambda)\in\R^{T+1}$, $\Widots(\lambda)\in\R^{T\times (T+1)}$ and $\uidots(\lambda)\in\R^{T+1}$. 
Then, let $\hat \beta (\lambda)$ denote the OLS estimator for $\beta$ in model \eqref{main demeaned model old} from the subsample of $\lfloor n \lambda \rfloor$ individuals, that is
 \begin{eqnarray*}
     \hat \beta(\lambda) &=& (\hat\Gamma(\lambda))^{-1} \frac {1}{\lfloor n \lambda \rfloor}
   \sum_{i\in\calS_{n,\lambda}} \ddot{W}_i(\lambda) \ddot{Y}_i(\lambda) 
   =\beta +  (\hat\Gamma(\lambda))^{-1} \frac {1}{\lfloor n \lambda \rfloor} \sum_{i\in\calS_{n,\lambda}} \ddot{W}_i(\lambda) \ddot{u}_i(\lambda),
 \end{eqnarray*}
 where the matrix $\Gamma(\lambda)$ is defined by  $ \hat \Gamma(\lambda) = \frac{1}{\lfloor n \lambda \rfloor} \sum_{i\in\calS_{n,\lambda}} \ddot{W}_i(\lambda) \ddot{W}_i(\lambda)'$. Now, assume that $\Gamma=\plim_{n\to\infty}\frac{1}{n}\sumi \ddot{W}_i \ddot{W}_i'$ is non-singular, where $\ddot{W}_i=\ddot{W}_i(1)$ and other variables are defined analogously. Under random sampling across individuals (see, e.g., \citealp{callaway2020difference}) and assuming that sufficient moments exist, it then follows that
 $$ \sup_{\lambda \in [\varepsilon,1]} \| \hat \Gamma(\lambda) - \Gamma \| = o_\mathbb{P}(1) \mbox{ as } n\to\infty,$$
 and 
 $$
 \sqrt{n} (\hat \beta(\lambda) - \beta) = \Gamma^{-1} \frac {\sqrt{n}}{\lfloor n \lambda \rfloor} \sum_{i\in\calS_{n,\lambda}} \ddot{W}_i(\lambda)   \ddot{u}_i(\lambda)   + o_\mathbb{P}(1)
 $$
 uniformly with respect to $\lambda  \in [\varepsilon,1]$. Consequently,  we obtain  from the Cramer-Wold device and Theorem 2.12.1 in
 \cite{vandervaart1996} that
 \begin{equation*}
   \big \{ \sqrt{n} (\hat \beta (\lambda) - \beta)\big \}_{\lambda \in [\varepsilon,1]} \rightsquigarrow \Big \{ \frac {V^{1/2}}{\lambda} \vec{\mathbb{B} } (\lambda) \Big \}_{\lambda \in [\varepsilon,1]}
 \end{equation*}
 where $ \vec{\mathbb{B} } $ is a $T$-dimensional vector of independent Brownian motions, $V=\Gamma^{-1}W\Gamma^{-1}$ with
\[
W=\plim_{n\to\infty}\frac{1}{n}\sumi \ddot{W}_i\ddot{u}_i\ddot{u}_i'\ddot{W}_i',
\]
and the symbol $\rightsquigarrow$ means weak convergence in the space $(\ell^\infty[\varepsilon,1])^{T}$ of all $T$-dimensional bounded functions on the interval $[\varepsilon,1]$.

\section{Mathematical proofs}\label{appendix mathematical proofs}

 \subsection{Properties of the test \eqref{testt=2}}
 For sufficiently large sample sizes the quantile $f_\alpha := \mathcal{Q}_{\mathrm{N}_F(\delta,\hat{\Sigma}_{11}/n)}(\alpha)$ satisfies
 \begin{equation}\label{ha1}
  \alpha = \mathbb{P} \big (| \mathrm{N} (\delta, \hat \Sigma_{11}/n) | \leq \mathcal{Q}_{\mathrm{N}_F(\delta,\hat{\Sigma}_{11}/n)}(\alpha) \big )  =\Phi \Big( \frac {f_\alpha - \delta}{\Sigma_{11}} \Big) - \Phi \Big( \frac {- f_\alpha -  \delta}{\Sigma_{11}} \Big)+\mathrm{O}\left(\frac{1}{\sqrt{n}}\right)
 \end{equation}
 where $\Phi$ is the cdf of the standard normal distribution. Consequently, we obtain for the probability of rejection
 \begin{equation}\label{ha2}
 \mathbb{P}_{\beta_1} (|\hat \beta_1| \leq f_\alpha ) \approx \Phi \Big( \frac {f_\alpha - \beta_1}{\Sigma_{11}} \Big) - \Phi \Big( \frac {- f_\alpha - \beta_1}{\Sigma_{11}} \Big) .
 \end{equation}
 It is well known that the right-hand side of \eqref{ha2} (with the quantile $f_\alpha$ defined by \eqref{ha1}) is the power function of the uniformly most powerful unbiased test  (see Example 1.1 in \citealp{romano2005}).

  \subsection{Proof of Theorem \ref{thmneu}}   
  
  The proof follows essentially by the same arguments as given in \cite{detmolvolbre2015}, and, for the sake of brevity, we only explain why this is the case (also see the discussion below in Section \ref{a2}, where a sequential version of the result is derived). By Assumption \ref{ass-sequential}, $ \sqrt{n} (\hat \beta     - \beta  ) $ 
    has an asymptotic $T$-dimensional centered normal distribution.		
    We denote the corresponding asymptotic covariance matrix by $\Sigma = (\sigma_{ij})_{i,j=1, \ldots T}
$.
Now we interpret all vectors as stochastic processes on the set ${\cal X} =\{1, \ldots , T \}$ and rewrite the weak convergence of the vector $\hat \beta  =(\hat \beta_1 , \ldots , \hat \beta_{T} )'$
as 
\begin{align} \label{hol11}
  \{ \sqrt{n} (\hat \beta_x    - \beta_x )  \}_{x\in {\cal X}  } 
\rightsquigarrow  \{ \mathbb{G} (x)   \}_{x\in {\cal X}  } ,
    \end{align}
   where $\{ \mathbb{G} (x)   \}_{x \in {\cal X}} $ is a centered 
   Gaussian process on ${\cal X} =\{1, \ldots , T \}$ with covariance structure $\text{Cov} ( \mathbb{G} (x), \mathbb{G} (y) ) = \sigma_{xy}$ ($x,y \in {\cal X}$). Note that \eqref{hol11} is the analog of equation (A.7) in \cite{detmolvolbre2015}, and it follows by exactly the same arguments  as stated in this paper that 
   \begin{align} \label{hol12}
  \sqrt{n} \big ( \| \hat \beta   \|_\infty    - \| \beta   \|_\infty \big ) \rightarrow \max  \big\{ \max_{x\in {\cal E}^+} \mathbb{G} (x)
  , \max_{x\in {\cal E}^-} -\mathbb{G} (x) \big \}  ~, 
\end{align}
provided that $\| \hat \beta   \|_\infty  >0 $, 
  where  the sets $ {\cal E}^+$ and $ {\cal E}^-$ are defined by 
\begin{align*}
  {\cal E}^+ &= \{ \ell = 1, \ldots , T :~ \beta_\ell = \|\beta \|_\infty  \} ~,\\
   {\cal E}^-&= \{ \ell = 1, \ldots , T :~ \beta_\ell = -\|\beta \|_\infty  \}~,
\end{align*}
respectively.
Note that $ {\cal E}^-\cup  {\cal E}^+ =  {\cal E}$, where $ {\cal E}$ is defined in \eqref{hd11}, and that 
\eqref{hol11} is the analog of Theorem 3 in \cite{detmolvolbre2015}. Moreover, if   $ \hat \beta^{*} = ( \hat  \beta_1^*,  \ldots , \hat \beta_{T}^*)' $  denotes the  estimate from the bootstrap sample, we obtain an analog of the weak convergence in  \eqref{hol11}, that is
\begin{align}  \label{hol13}
   \{ \sqrt{n} (\hat \beta_x^*     - \hat{\hat \beta}_x  )  \}_{x\in {\cal X}  }  
& \rightsquigarrow  \{ \mathbb{G} (x)   \}_{x\in {\cal X}  } 
\end{align}
conditional on the sample $(\ddot{W}_1, \ddot{Y}_1),\ldots,(\ddot{W}_{n},\ddot{Y}_n)$. Note that this statement corresponds to the statement (A.25) in \cite{detmolvolbre2015}. 
Now the statements (A.7) and (A.25) and their Theorem 3 are the main ingredients for the proof of Theorem 5 in \cite{detmolvolbre2015}. In the present context these statements can be replaced by  \eqref{hol11},  \eqref{hol13} and \eqref{hol12}, respectively, and a careful inspection of the arguments given in \cite{detmolvolbre2015} shows that  Theorem \ref{thmneu} holds
(the arguments even simplify substantially  as in our case the index set ${\cal X} $ of the processes  is finite).

 \subsection{Proof of Theorem \ref{thm0}} \label{a2} 
Let Assumption \ref{ass-sequential} hold. In the case $\beta  =0$ the result in Theorem \ref{thm0} now follows directly from the continuous mapping theorem. On the other hand, if  $\beta  \not =0$, it follows that
\begin{eqnarray*}
H_n (\lambda)  & =&
\sqrt{n}  \big ( \| \hat \beta  (\lambda)\|^2 - \| \beta    \|^2 \big)  \\
&=& \sqrt{n} \{   \| \hat \beta  (\lambda) - \beta \|^2 +
2(\hat \beta  (\lambda)-\beta )' \beta  \\
&=&   2 \sqrt{n}(\hat \beta  (\lambda) - \beta  )'   \beta    + o_{\mathbb{P}}(1)
\end{eqnarray*}
uniformly with respect to $\lambda \in [\varepsilon,1]$, and a further application of the  continuous mapping theorem yields
$$
\big \{ H_n (\lambda) \big \}_{\lambda \in [\varepsilon,1]} \rightsquigarrow \Big \{   2\beta ' D  \frac { \vec{\mathbb{B} } (\lambda)}{\lambda} \Big \}_{\lambda \in [\varepsilon,1]}
$$
in $\ell^\infty([\varepsilon,1])$.
It is easy to see that  for $\beta \neq 0$ the process on the right-hand side equals in distribution
$$
\Big \{ \Delta  (\beta ) \frac {\mathbb{B}_1 (\lambda)}{\lambda} \Big \}_{\lambda \in [\varepsilon,1]}
$$
where $\mathbb{B}_1$ is a one-dimensional Brownian motion and 
\begin{align}
    \label{hol21}
   \Delta  (\beta )  = 4\beta' D  D'  \beta
\end{align}
is a positive constant. Recalling the definition of the statistic $\hat M_n$ in \eqref{mt}  and
 a further application of the continuous mapping theorem shows that
\begin{align*}
\hat M_n &= \frac{\hat{\beta}_{RMS}^2(1)-  \beta_{RMS}^2}{\hat{V}_n}  
\\&
=\frac {\| \hat \beta (1)\|^2 - \| \beta    \|^2}{\big( \int^1_\varepsilon ( \|  \hat \beta  (\lambda)\|^2 - \| \hat \beta  (1) \|^2  )^2   \nu (d \lambda) \big)^{1/2}}  \\
    &= \frac {H_n (1)}{\big( \int^1_\varepsilon (H_n(\lambda) - H_n(1))^2 \nu (d \lambda) \big)^{1/2}} 
    \\ & 
    \overset{d}{\to} \mathbb{W}=
 \frac {\mathbb{B}_1 (1)}{\big( \int^1_\varepsilon (\mathbb{B}_1(\lambda)/\lambda - \mathbb{B}_1(1))^2 \nu (d \lambda)  \big)^{1/2}},
\end{align*}
which proves the assertion.

\subsection{Proof of Theorem \ref{thm1}} Observing the definition of $\hat M_T$  in \eqref{mt} we obtain
\begin{equation*}
 \mathbb{P}_{ {\beta}} \Big(\hat{\beta}_{RMS}^2 < \zeta^2 + \mathcal{Q}_{\mathbb{W}}(\alpha) \hat{V}_n
 \Big) 
 = \mathbb{P}_{ {\beta}}  \Big( \hat M_n  < \frac{\zeta^2 - {\beta}_{RMS}^2}{ \hat V_n} + Q_{\mathbb{W}}(\alpha) \Big).
\end{equation*}
It follows from the proof of Theorem \ref{thm0}  that $\hat V_n= O_{\mathbb{P}}(1/\sqrt{n})$. Consequently, if  $\beta_{RMS}^2 > 0 $,  the  assertion   follows by a simple calculation considering the three cases separately. On the other hand, if $\beta_{RMS} = 0$, the proof of Theorem \ref{thm0}  also shows that $\| \hat \beta  (1)\|^2 = O_{\mathbb{P}}(\frac {1}{n})$ and the assertion follows from the weak convergence \eqref{hol31} in Theorem \ref{thm0}.

\clearpage
\section{Simulation results}
\newcommand{\ra}[1]{\renewcommand{\arraystretch}{#1}}

\begin{table}[h!]
\centering
\ra{1.2}
\begin{tabular}{@{}lcccccccccc@{}}
\toprule
& \multicolumn{4}{c}{$n = 100$} & \phantom{abc} & \multicolumn{4}{c}{$n = 1000$} \\
\cmidrule{2-5} \cmidrule{7-10}
$\mathrm{Test}$ & $T=1$ & $T=4$ & $T=8$ & $T=12$ & & $T=1$ & $T=4$ & $T=8$ & $T=12$ \\
\midrule
\eqref{test statistic BOT}& 0.0546 & 0.0043& 0.0020 & 0.0004 &&0.0470 & 0.0046 & 0.0013 & 0.0009 \\
\eqref{test statistic bootstrap}& 0.0517 & 0.0134& 0.0092 & 0.0073 &&0.0477 & 0.0137 & 0.0103 & 0.0069 \\
\eqref{test statistic bootstrap}$_\text{cluster}$& 0.0520 & 0.0114& 0.0088 & 0.0079 &&0.0483 & 0.0116 & 0.0086 & 0.0065 \\
\eqref{test statistic mean beta}& 0.0546 & 0.0528& 0.0517 & 0.0537 &&0.0470 & 0.0493 & 0.0532 & 0.0491 \\
\eqref{test statistic rms}& 0.0727 & 0.0663& 0.0737 & 0.0801 &&0.0469 & 0.0450 & 0.0487 & 0.0500 \\
\bottomrule
\end{tabular}
\caption{Rejection frequencies for $\beta_t=1$, $t=1,\ldots,T$ with equivalence threshold 1 at nominal level of significance $\alpha=5\%$.}
\label{rejection frequencies ar3 1}
\end{table}

\begin{table}[h!]\centering
\ra{1.2}
\begin{tabular}{@{}lrrrrcrrrr@{}}\toprule
& \multicolumn{4}{c}{$n  = 100$} & \phantom{abc}& \multicolumn{4}{c}{$n  = 1000$} \\
\cmidrule{2-5} \cmidrule{7-10}
$\mathrm{Test}$& $T=1$ & $T=4$ & $T=8$ &$T=12$ && $T=1$ & $T=4$ & $T=8$ &$T=12$\\
 \midrule
\eqref{test statistic BOT} & 0.0554 & 0.0394 & 0.0089 & 0.0013 && 0.0507 & 0.0512 & 0.0498 & 0.0502\\
\eqref{test statistic bootstrap}& 0.0526 & 0.0998 & 0.1283 & 0.1417 && 0.0516 & 0.0866 & 0.1038 & 0.1126\\
\eqref{test statistic bootstrap}$_\text{cluster}$& 0.0544 & 0.0839 & 0.1188 & 0.1316 && 0.0520 & 0.0513 & 0.0505 & 0.0505\\
\eqref{test statistic mean beta} & 0.0554 & 0.9690 & 0.9709 & 0.9446 && 0.0507 & 1.0000 & 1.0000 & 1.0000\\
\eqref{test statistic rms}& 0.0761 & 0.7331 & 0.7590 & 0.6957 && 0.0482 & 0.9998 & 1.0000 & 1.0000\\
\bottomrule
\end{tabular}
\caption{Rejection frequencies for $\beta_1=1$ and $\beta_l=0$, $l=2,...,T$ with equivalence threshold 1 at nominal level of significance $\alpha=5\%$.  }
\label{rejection frequencies ar3 2}
\end{table}

\begin{table}[h!]\centering
\ra{1.2}
\begin{tabular}{@{}lrrrrcrrrr@{}}\toprule
& \multicolumn{4}{c}{$\beta_l = 0.8,\, l=1,...,T$} & \phantom{abc}& \multicolumn{4}{c}{$\beta_l = 0.9,\, l=1,...,T$} \\
\cmidrule{2-5} \cmidrule{7-10}
$\mathrm{Test}$& $T=1$ & $T=4$ & $T=8$ &$T=12$ && $T=1$ & $T=4$ & $T=8$ &$T=12$\\
 \midrule
\eqref{test statistic BOT}& 0.8807 & 0.5460 & 0.3132 & 0.2017 && 0.4042 & 0.1099 & 0.0456 & 0.0246\\
\eqref{test statistic bootstrap}& 0.8805 & 0.6827 & 0.5489 & 0.4507 && 0.4058 & 0.2034 & 0.1401 & 0.1060\\
\eqref{test statistic bootstrap}$_\text{cluster}$& 0.8809 & 0.6397 & 0.4838 & 0.3965 && 0.4068 & 0.1807 & 0.1238 & 0.0984\\
\eqref{test statistic mean beta}& 0.8807 & 0.8958 & 0.8197 & 0.7500 && 0.4042 & 0.4276 & 0.3542 & 0.3112\\
\eqref{test statistic rms}& 0.7000 & 0.7146 & 0.6268 & 0.5675 && 0.2935 & 0.3006 & 0.2575 & 0.2311\\
\bottomrule
\end{tabular}
\caption{Rejection frequencies for $n=1000$ with equivalence threshold 1 at nominal level of significance $\alpha=5\%$.}
\label{power all ar3}
\end{table}

\begin{table}[h!]\centering
\ra{1.2}
\begin{tabular}{@{}lrrrrcrrrr@{}}\toprule
& \multicolumn{4}{c}{$\beta_1 = 0.8$, $\beta_l=0$, $l=2,...,T$} & \phantom{abc}& \multicolumn{4}{c}{$\beta_1 = 0.9$, $\beta_l=0$, $l=2,...,T$} \\
\cmidrule{2-5} \cmidrule{7-10}
$\mathrm{Test}$& $T=1$ & $T=4$ & $T=8$ &$T=12$ && $T=1$ & $T=4$ & $T=8$ &$T=12$\\
 \midrule
\eqref{test statistic BOT}& 0.8776 & 0.6803 & 0.5274 &  0.4479 && 0.4103 & 0.2768 & 0.2161 & 0.1863\\
\eqref{test statistic bootstrap}& 0.8763 & 0.7746 & 0.6768  & 0.6207 && 0.4108 & 0.3769 & 0.3472 & 0.3191\\
\eqref{test statistic bootstrap}$_\text{cluster}$& 0.8771 & 0.6804 & 0.5299  & 0.4488 && 0.4103 & 0.2794 & 0.2172 & 0.1878\\
\eqref{test statistic mean beta}& 0.8776 & 1.0000 & 1.0000  & 1.0000 && 0.4103 & 1.0000 & 1.0000 & 1.0000\\
\eqref{test statistic rms}& 0.6920 & 1.0000 & 1.0000 &  1.0000 && 0.3003 & 1.0000 & 1.0000 & 1.0000\\
\bottomrule
\end{tabular}
\caption{Rejection frequencies for $n=1000$ with equivalence threshold 1 at nominal level of significance $\alpha=5\%$. }
\label{power single ar3}
\end{table}

\begin{table}[h]\centering
\begin{tabular}{@{}rrrrrcrrrr@{}}\toprule
& \multicolumn{4}{c}{$n  = 100$} & \phantom{abc}& \multicolumn{4}{c}{$n  = 1000$} \\
\cmidrule{2-5} \cmidrule{7-10}
& $T=1$ & $T=4$ & $T=8$ &$T=12$ && $T=1$ & $T=4$ & $T=8$ &$T=12$\\
 \midrule
$\#insig/M$ & 0.9512 & 0.8340&  0.7204& 0.6468&& 0.9516 & 0.8472& 0.7392 & 0.6800 \\
$\hat{\pi}_{ATT}$& 0.0019 & -0.0066 & -0.0036 & 0.0031 &&  -0.0001 & 0.0002 & 0.0001 & -0.0008\\
$\mathrm{CI}^{\hat{\pi}_{ATT}}$& 0.9532 & 0.9512  & 0.9460  & 0.9496 &&  0.9528  & 0.9496 & 0.9480 & 0.9564 \\
  $\delta^*_{IU}$& 0.6300 & 0.8491  & 0.9265  & 0.9626 &&  0.2018  & 0.2673 & 0.2930 & 0.3045 \\
$\delta^*_{Boot}$& 0.6426 & 0.6875  & 0.7020  & 0.7017 &&  0.2029  & 0.2128 & 0.2194 & 0.2208 \\
$\delta^*_{c.Boot}$& 0.6401 & 0.6915  & 0.7040  & 0.7049 &&  0.2028  & 0.2128 & 0.2195 & 0.2210 \\
$\tau^*$& 0.6295 & 0.5029  & 0.4808  & 0.4662 &&  0.2017  & 0.1570 & 0.1527 & 0.1477 \\
$\zeta^*$& 0.7132 & 0.6961  & 0.6884  & 0.6881 &&  0.2247  & 0.2166 & 0.2142 & 0.2153\\
\bottomrule
\end{tabular}
\caption{Treatment effect estimation ($\pi_{ATT}=0$) and smallest equivalence thresholds under the PTA at nominal level of significance $\alpha=5\%$. }
\label{simulation pta homoskedastic}
\end{table}

\begin{table}[h]\centering
\ra{1.2}
\begin{tabular}{@{}rrrrrcrrrr@{}}\toprule
& \multicolumn{4}{c}{$n  = 100$} & \phantom{abc}& \multicolumn{4}{c}{$n  = 1000$} \\
\cmidrule{2-5} \cmidrule{7-10}
& $T=2$ & $T=4$ & $T=8$ &$T=12$ && $T=2$ & $T=4$ & $T=8$ &$T=12$\\
 \midrule
$\#insig/M$&0.8256 & 0.6668 & 0.5532 &0.4456 && 0.1560 &  0.1016& 0.1816& 0.0772\\
$\hat{\pi}_{ATT}$& -0.2501 & -0.2528 & -0.2574 & -0.2580 && -0.2495 & -0.2495 & -0.2450 & -0.2499\\
$\mathrm{CI}^{\hat{\pi}_{ATT}}$& 0.8276  & 0.8060  & 0.6820  & 0.8916  && 0.1436  & 0.2044  & 0.3408  & 0.2868 \\
  $\delta^*_{IU}$& 0.7035  & 0.9763  & 1.3027  & 1.3185  && 0.3916  & 0.4613  & 0.5188  & 0.5528 \\
$\delta^*_{Boot}$& 0.7173  & 0.8000  & 1.0382  & 0.9474  && 0.3937  & 0.4194  & 0.4438  & 0.4657 \\
$\delta^*_{c.Boot}$& 0.7141  & 0.8622  & 1.1419  & 1.0820  && 0.3936  & 0.4456  & 0.4786  & 0.5031 \\
$\tau^*$& 0.7032  & 0.7018  & 0.8940  & 0.8073  && 0.3916  & 0.3838  & 0.3979  & 0.4082 \\
$\zeta^*$& 0.7619  & 0.8266  & 1.0326  & 0.9845  && 0.4025  & 0.4022  & 0.4258  & 0.4321\\
\bottomrule
\end{tabular}
\caption{Treatment effect estimation ($\pi_{ATT}=0$) and smallest equivalence thresholds under a violation of the PTA due to a temporary group-specific shock with mean $0.25$ at nominal level of significance $\alpha=5\%$. }
\label{simulation ashenfelter 025}
\end{table}

\begin{table}[h]\centering
\begin{tabular}{@{}rrrrrcrrrr@{}}\toprule
& \multicolumn{4}{c}{$n  = 100$} & \phantom{abc}& \multicolumn{4}{c}{$n  = 1000$} \\
\cmidrule{2-5} \cmidrule{7-10}
& $T=2$ & $T=4$ & $T=8$ &$T=12$ && $T=2$ & $T=4$ & $T=8$ &$T=12$\\
 \midrule
$\#insig/M$&0.5240 & 0.3468  & 0.3080 &0.2928 && 0.0000 &  0.0000& 0.0076& 0.0000\\
$\hat{\pi}_{ATT}$& -0.5020 & -0.5027 & -0.4971 & -0.4942 &&  -0.4973 & -0.5008 & -0.5005 & -0.5017\\
$\mathrm{CI}^{\hat{\pi}_{ATT}}$& 0.5040  & 0.5384  & 0.5500  & 0.6760  &&  0.0000  & 0.0004  & 0.0168  & 0.0000 \\
 $\delta^*_{IU}$& 0.9245  & 1.1822  & 1.4257  & 1.4784  &&  0.6367  & 0.7121  & 0.7648  & 0.8013 \\
$\delta^*_{Boot}$& 0.9377  & 1.0416  & 1.1826  & 1.1626  &&  0.6371  & 0.6718  & 0.6980  & 0.7182 \\
$\delta^*_{c.Boot}$& 0.9338  & 1.1149  & 1.2897  & 1.3010  &&  0.6371  & 0.6981  & 0.7350  & 0.7566 \\
$\tau^*$& 0.9244  & 0.9236  & 1.0280  & 0.9966  &&  0.6367  & 0.6357  & 0.6471  & 0.6587 \\
$\zeta^*$& 0.9527  & 1.0077  & 1.1419  & 1.1178  &&  0.6661  & 0.6722  & 0.6820  & 0.6932 \\
\bottomrule
\end{tabular}
\caption{Treatment effect estimation ($\pi_{ATT}=0$) and smallest equivalence thresholds under a violation of the PTA due to a temporary group-specific shock with mean $0.5$ at nominal level of significance $\alpha=5\%$.}
\label{simulation ashenfelter 05}
\end{table}

\begin{table}[h]\centering
\begin{tabular}{@{}rrrrrcrrrr@{}}\toprule
& \multicolumn{4}{c}{$n  = 100$} & \phantom{abc}& \multicolumn{4}{c}{$n  = 1000$} \\
\cmidrule{2-5} \cmidrule{7-10}
& $T=2$ & $T=4$ & $T=8$ &$T=12$ && $T=2$ & $T=4$ & $T=8$ &$T=12$\\
 \midrule
$\#insig/M$& 0.9280&  0.8156  &  0.6568 & 0.4848 &&  0.9456 &  0.6736 &0.1876  & 0.0144\\
$\betahatTplusone$& 0.0304 & 0.0240 & 0.0337 & 0.0208 && 0.0261 & 0.0230 & 0.0238 & 0.0253\\
$\mathrm{CI}^{\hat{\pi}_{ATT}}$& 0.9432 & 0.9456 & 0.9380 & 0.9516 && 0.9348 & 0.9456 & 0.9416 & 0.9452\\
$\delta^*_{IU} $& 0.6474 & 0.8596 & 0.9551 & 1.0392 && 0.2039 & 0.2992 & 0.3892 & 0.4859\\
$\delta^*_{Boot} $& 0.6604 & 0.7060 & 0.7454 & 0.8189 && 0.2052 & 0.2589 & 0.3538 & 0.4590\\
$\delta^*_{c.Boot} $& 0.6580 & 0.7101 & 0.7470 & 0.8221 && 0.2048 & 0.2586 & 0.3538 & 0.4587\\
$\tau^*$& 0.6470 & 0.5103 & 0.4978 & 0.5203 && 0.2038 & 0.1853 & 0.2229 & 0.2696\\
$\zeta^*$& 0.7396 & 0.7055 & 0.7048 & 0.7144 && 0.2282 & 0.2342 & 0.2643 & 0.3122\\
\bottomrule
\end{tabular}
\caption{Treatment effect estimation ($\pi_{ATT}=0$) and smallest equivalence thresholds under a violation of the PTA due to a time trend with slope $0.025$ at nominal level of significance $\alpha=5\%$.}
\label{time trend 0025 new homoskedastic}
\end{table}

\end{document}